\documentclass[aps,prb,twocolumn,floatfix,superscriptaddress]{revtex4-1}
\usepackage{comment}
\usepackage{graphicx}
\usepackage{amsmath}
\usepackage{amssymb}
\usepackage{slashed}
\usepackage{braket}
\usepackage{color}
\usepackage[normalem]{ulem}
\usepackage[dvipsnames]{xcolor}

\newcommand{\non}{\nonumber}
\newcommand{\bea}{\begin{eqnarray}}
	\newcommand{\eea}{\end{eqnarray}}
\newcommand{\beq}{\begin{equation}}
	\newcommand{\eeq}{\end{equation}}

\begin{document}

\title{ Physics of the Inverted Harmonic Oscillator: From the lowest Landau level to event horizons}

\author{Varsha Subramanyan} 
\thanks{These authors contributed equally.}
\affiliation{Dept. of Physics, University of Illinois, Urbana-Champaign,US}
\author{Suraj S. Hegde}
\thanks{These authors contributed equally.}
\affiliation{Max-Planck-Instiut f{\"u}r Physik komplexer systeme, Dresden, Germany}
\author{Smitha Vishveshwara}
\email{smivish@illinois.edu}
\author{Barry Bradlyn}
\email{bbradlyn@illinois.edu}
\affiliation{Dept. of Physics, University of Illinois, Urbana-Champaign,US}

\begin{abstract}
In this work, we present the inverted harmonic oscillator (IHO) Hamiltonian as a paradigm to understand the quantum mechanics of scattering and time-decay in a diverse set of physical systems. As one of the generators of area preserving transformations, the IHO Hamiltonian can be studied as a dilatation generator, squeeze generator, a Lorentz boost generator, or a scattering potential. In establishing these different forms, we demonstrate the physics of the IHO that underlies phenomena as disparate as the Hawking-Unruh effect and scattering in the lowest Landau level (LLL) in quantum Hall systems. We derive the emergence of the IHO Hamiltonian in the LLL in a gauge invariant way and show its exact parallels with the Rindler Hamiltonian that describes quantum mechanics near event horizons. This approach of studying distinct physical systems with symmetries described by isomorphic Lie algebras through the emergent IHO Hamiltonian enables us to reinterpret geometric response in the lowest Landau level in terms of relativistic effects such as Wigner rotation. Further, the analytic scattering matrix of the IHO points to the existence of quasinormal modes (QNMs) in the spectrum, which have quantized time-decay rates. We present a way to access these QNMs through wave packet scattering, thus proposing a novel effect in quantum Hall point contact geometries that parallels those found in black holes. 

\end{abstract}

\maketitle

\vspace{5ex}

 \tableofcontents

\section{Introduction}\label{Intro}

In the past several years, physics from the microscopic quantum scale to astronomical scales has enjoyed a surge of advances building on foundational work. As one of many instances, several theoretical and experimental efforts towards detecting signatures of quasiparticle fractional statistics in mesoscopic quantum Hall phases  have recently yielded tremendous success. Quantum Hall interferometer and  beam-splitter experiments, have at last begun to reveal unequivocal signatures\cite{Bartolomei20,Nakamura20} of such anyonic statistics. On the far extreme scale, black holes and their gravitational-wave signatures that had remained undetected for nearly half a century have finally been observed: Recent experiments with LIGO have decisively provided empirical evidence of black hole mergers again through gravitational wave interferometry\cite{LIGO1}. 

Although seemingly disjoint areas of study, recent progress has shown that there are ideas and techniques common to both condensed matter and black hole physics\cite{Franz18,Hartnoll,Hegde19}. Today, we are witnessing a rapid expansion in this cross—fertilization between sub-fields, giving rise to tremendous insights and far-reaching predictions. Investigations in quantum gravity and condensed matter have married concepts from general relativity, quantum field theory and quantum information. For instance, attempts to address the problem of understanding quantum effects near an event horizon have been the seeds of many modern developments such as the SYK model and random unitary circuits, bringing forth key concepts such as chaos and complexity. Concepts that have historically emerged in one realm have found their identity as intrinsic structures in otherwise unrelated systems. This is exemplified in a plethora of concepts that originated in high energy physics but have since found their identity in quantum condensed matter---the Dirac equation, monopoles, skyrmions, Majorana fermions, and more. The power of such parallels is evident in instances found through the ages, such as in the Higgs-Anderson mechanism. Furthermore, symmetry provides guiding principles in finding commonalities in seemingly disparate systems, be it conservation laws derived from Noether’s theorem, universality in symmetry broken phases, or symmetry protected topological phases. Model Hamiltonians oftentimes serve as the embodiment of symmetry manifestations. Highly complex behavior can at least in part be described in terms of a very simple model, offering a mine of experimentally verifiable information; the simple harmonic oscillator (SHO) is the paragon for such models.

In our work, we offer common ground for various threads in the astrophysical and condensed matter realmsby way of a relatively overlooked unifying model - the inverted harmonic oscillator (IHO). Sister to the SHO, the IHO has remarkable properties in its own right that make their way across disciplines\cite{Barton86,Maldacena05,Hegde19,Sierra,Bhaduri,Bhattacharyya20,Betzios16,Dalui19,Friess,Gentilini,Morita19}. As with the SHO, the quantum treatment of the IHO is completely solvable. While the SHO effectively models deviations from a stable equilibrium point, the IHO acts as an accurate approximation for decay from an unstable equilibrium,  and comes with a whole mathematical machinery for treating scattering and decaying states. The IHO can at once be perceived as a generator of squeezing common to quantum optics, as a dilatation generator for scaling behaviour, and as a quantum mechanical scattering barrier. The IHO is also related to relativistic Lorentz boosts as we show in this work. Another key feature of the IHO is the presence of quasinormal modes--temporally decaying modes having quantized decay rates--a unique manifestation of quantization. The IHO thus acts as an excellent prototype for describing situations involving  tunneling and decay, as for instance pointed out decades ago for nuclear processes. Here, we show that this simple but rich model naturally occurs in the two very different settings of quantum Hall lowest Landau level physics and physics at black hole event horizons. Phenomena in these settings have an equivalence dictated by the IHO Hilbert space and the time evolution governed by its Hamiltonian. We draw attention to how, remarkably, processes as different as Hawking-Unruh radiation from black holes and quasiparticle tunneling in quantum Hall point contacts stem from the same underlying IHO physics.

As one realm of focus in our work, the quantum Hall system consists of a two-dimensional electron gas subject to high magnetic fields, typically on the mesoscopic scale in low-temperature lab settings\cite{Girvin,Tong}. It is hailed for supporting persistent edge currents and quantized conductance comprised purely of fundamental constants. Topological aspects of the quantum Hall fluid lie behind conductance quantization. In the case of fractionally filled states, they give rise to anyonic quasi-particles having fractional charge which has been measured in point contact geometries. The IHO makes its way into this system in two ways, both relying on the non-commutative nature of the lowest Landau level. Shear potentials applied on the system as well as saddle potentials characteristic  point contacts and disordered landscapes, can both be treated in terms of one-dimensional quantum mechanical IHOs. 

As the other realm of focus, black holes are one of the most intriguing astrophysical objects. They are the simplest macroscopic objects in nature in that they are described purely by their mass, angular momentum, and charge\citep{Chandrasekhar}. Their fundamental description, at least classically, is purely geometrical and yet they exhibit a plethora of features such as singularities, one-way propagation, and quasinormal modes\citep{MTW}. In this setting too, the IHO naturally occurs in two guises. At the classical level, in the simplest case of a Schwarzschild black hole, spherical symmetry gives rise to a one-dimensional scattering potential along the radial direction beyond the event horizon\cite{Schutz85}. Further, the process of Hawking radiation entails quantum fluctuations across the event horizon that are intimately tied to IHO-based Rindler time evolution\citep{Betzios16}. A black hole is the key phenomenological entity in nature that forms the ground for interplay between quantum mechanics and gravity. Here we draw attention to the commonality between tunneling across the saddle potential in the quantum Hall system and Hawking-Unruh radiation resulting from the emergence of thermal bath for a uniformly accelerating observer in Minkowski space-time stem from Rindler time-evolution associated with the IHO. Quantum Hall point-contact tunneling conductance and the thermal form of radiation from black holes are therefore identical in formal structure. Another signature feature of IHO physics are QNMs, which in the context of black holes have also played a key role in the unequivocal detection of black holes through gravitational waves\cite{Vishveshwara, Konoplaya11}. As we pointed out in recent work \cite{Hegde19},  detecting quasinormal mode decay through pulsed high-frequency measurement in point contacts would constitute a new observation in this mesoscopic realm. 

In what follows, we will explore in detail the properties of the IHO and its role as a conceptual glue between these diverse areas. Given the comprehensive nature of the work, we begin with a summary of our main results and provide a road map for reading the manuscript. In the subsequent Section \ref{Dynamics}, we chart out the instances mentioned above in which the IHO appears in the quantum Hall system. In Sec. \ref{IHO}, following a survey of the IHO from different perspectives, we present its scattering properties, including a discussion of the scattering matrix. We then bring focus to quasinormal modes in Sec. \ref{QNM}, and show how these resonances persist in realistic potentials. We will pay particular attention to observable signatures of QNMs. In Sec. \ref{RindHam}, we lay out the elegant machinery behind the Rindler Hamiltonian underlying the IHO and its time evolution, and present the manner in which it gives rise to Hawking-Unruh physics. In Sec. \ref{Wigner}, we move on to symmetry considerations, showing that the parallels between phenomena can be framed in terms of the underlying Lie-algebra isomorphisms. We show how an effect of Lorentz kinematics such as the Wigner rotation could be captured in a quantum Hall setting. In Sec. \ref{WaveEq}, we show how IHO QNMs are related to their black hole counterparts through effective scattering problem in the wave equation of fields in black hole spacetime. Finally, in Sec.\ref{Outlook} we present a roadmap to vast number of topics where the IHO is relevant and discuss various avenues closely tied to our work.

Before we embark on our exposition, we acknowledge the circumstances under which it came to shape: this work was brought to completion during the midst of the global Covid-19 pandemic, which created an exceptional set of challenges for the world at large. At the same time, the physics community was rocked with the loss of luminaries like Philip Anderson and Margaret Burbidge, whose pioneering contributions to the fields of condensed matter and astrophysics relate to the core of this present work. Pockets of light still persisted. On the scientific front most connected to this work, advances in quantum and astrophysics surged forward, building on foundational work across decades. It is a marvel that within months of each other, not one but two separate ideas for detecting anyons in quantum Hall systems were experimentally realized and reported in Refs.~\onlinecite{Bartolomei20,Nakamura20}. On the black hole front, this year, multi-messenger astronomy provided many new insights while also marking the 50th anniversary of the original prediction of black hole quasinormal modes by C.~V.~Vishveshwara. The year's Nobel recognition combined R.~Penrose's decades-old fundamental work on the existence of black holes with more recent discovery of a supermassive compact object at the center of galaxy by the groups of R.~Genzel and A.~Ghez. These highlights represent but an iota of the enduring science persevered by thousands of researchers across the globe. Our work serves as a tribute to these reminders that transformative ideas have the power to transcend challenges and tragedies.

\section{Summary of Main Results and structure of the paper.}
\label{Summary}

This work is partially a presentation of original results, and partially a perspective-review. We expand and elucidate the key concepts underlying the results presented by us in our short paper \cite{Hegde19} and clarify the distinction between similar looking scenarios.

The line of reasoning we have pursued here is to present the existence of an equivalence between the bare minimum quantum mechanical structures present in problems that are physically distinct at the level of phenomenology and experiments. The observables in these different settings would correspond to very different kind of measurements (as distinct as a thermal distribution of Hawking radiation and the conductance in quantum Hall). Yet these `expectation values' nevertheless come from the following key mathematical structures of quantum mechanics:1) The states/wavefunctions, the associated Hilbert space and its representation, 2) the algebra of operators acting on them (which correspond to physical quantities), and 3) The evolution with respect to a Hamiltonian and its symmetry (the group structure).
 The inquiry pursued here is to ask if, in two distinct physical settings, there exists an equivalence between the underlying quantum mechanical structures elucidated above, and if this underlies the `analogy'  between appearances at a phenomenological level. Quantum Hall physics under applied potentials and Hawking-Unruh effect are the two such phenomena under consideration in this paper. In this section, we shall summarise the exact points at which we have seen equivalence between these two phenomena and how this line of reasoning has led to exploration of new kind of experimental probes  and novel understanding of known quantum Hall physics.

The key results we present in this paper are the following:

\begin{itemize}
 
    \item We highlight the importance of the inverted Harmonic oscillator Hamiltonian as the key structure underlying the parallel between Hawking-Unruh effect and quantum Hall point contact geometry. The `Gibbs' thermal-like factor $e^{-\beta E}$ and the thermal-like distribution form $1/ (\exp(-\beta E)+1)$ appear as scattering amplitudes and tunneling probabilities across the IHO potential.(Sec.\ref{IHO}).  We put forth different `avatars' of the IHO as a scattering potential, generator of squeezed states, and as a dilatation generator.
    
    \item  We show that the emergence of thermal-like factors in the context of event horizons of black holes and the quantum Hall point contact set-up is rooted in the equivalence between the wavefunctions of the Rindler modes/Lorentz boost eigenmodes and the IHO eigensystem(Sec.\ref{BoostIHO}). In these mappings, the role of temperature is played by the strength of the point-contact potential in the quantum Hall setting and by the surface gravity in the black hole setting.   The Lorentz boost generator in fact takes the role of a Hamiltonian for quantum mechanical states near a space-time horizon. This Hamiltonian is called the 'Rindler Hamiltonian' and is a fundamental object in studying entanglement properties of horizons and in topological phases of condensed matter. We relate the Rindler Hamiltonian to the IHO. Thus, quantum mechanics in a relativistic setting  is made fully accessible in an experimentally viable set up of quantum Hall system under point-contacts.
    
    \item We provide a gauge-invariant derivation for the appearance of the IHO Hamiltonian in the lowest Landau level limit of a quantum Hall system that is under the influence of a saddle potential. We include this potential as a member of a class of quadratic potentials (electrostatic and those generated by strain) which on Landau level projection form the generators of the Lie-algebra of area-preserving deformations in two dimensions $\mathfrak{sl}(2,\mathbb{R})$. Stated another way,  they form the Lie algebra of linear canonical transformations $\mathfrak{sp}(2,\mathbb{R})$ that preserve the non-commutativity in the lowest Landau level.(Sec. \ref{Dynamics})

    \item We  present an important phenomena rooted in the physics of the IHO Hamiltonian and hitherto unexplored in the context of quantum Hall systems -  time-decaying states with quantized decay rates called `quasi-normal modes'. In general, quasinormal modes(QNM) are ubiquitous in scattering theory and appear as resonant modes. Here we provide a comprehensive description of these modes. We show how such states could be tapped through wave packet scattering in a quantum Hall setting. As a physical alternative to the unbounded potential of IHO, we present an analysis of the P\"{o}schll-Teller potential and compute quantities such as survival probability, which could be accessed through experiments (Sec. \ref{QNM}). These decaying modes are also known in the context of black hole physics. The quantized decay rates carry information on black hole parameters and  have proved as crucial signatures in recent detection of gravitational waves from black hole mergers through LIGO. 
    
    \item Finally, we point to a chain of Lie-algebra isomorphisms $\mathfrak{sl}(2,\mathbb{R}) \sim \mathfrak{sp}(2,\mathbb{R})\sim \mathfrak{su}(1,1) \sim \mathfrak{so}(2,1)$, where each of these algebras contains as its member the key structures we have discussed so far: $\mathfrak{sl}(2,\mathbb{R}) \sim \mathfrak{sp}(2,\mathbb{R})$(algebra of area-preserving deformations and linear canonical transformations) contains IHO  acting as a projected Hamiltonian in the lowest landau level, $\mathfrak{su}(1,1)$ contains IHO as a generator of squeezed-coherent states and finally $\mathfrak{so}(2,1)$ (algebra of Lorentz group) contains Lorentz boost generator. We suggest that these isomorphisms could underlie the equivalence between eigensystems of Lorentz boosts and the IHO; and the associated physical phenomena of Hawking-Unruh effect and scattering in lowest Landau levels.
    
    \item This analysis immediately leads us to explore other Lorentz kinematic effects such as Wigner rotation in the context of quantum Hall systems, by application of electrostatic and strain potentials.

    \end{itemize}
    
    Though the IHO Hamiltonian appears in both the contexts of Hawking-Unruh effect and quasinormal modes of black holes, we would like to clarify the difference in the physics of the two scenarios. In the context of Black hole QNMs, the potential barrier that appears in the scattering of fields in a black hole spacetime and which we approximate as an IHO lies outside the event horizon. The scattering could be in a purely classical scenario, as was done in the original work that proposed QNMs \cite{Vishveshwara}, without invoking any quantum mechanical degrees of freedom. On the other hand, the Hawking-Unruh effect is a purely quantum mechanical phenomena and the IHO appears in this context as a counterpart of the Rindler Hamiltonian which acts on the quantum mechanical states near the horizon. The Hawking-Unruh effect involves purely quantum mechanical effect across the horizon whereas QNMs in the context of scattering of fields against black holes are due to  classical scattering.
    
As mentioned earlier, part of this paper is a presentation of original work and part of it is review. The topics which we review here are mainly the physics of IHO, the Rindler Hamiltonian, its relation to Hawking-Unruh effect and its manifestations in entanglement aspects of condensed matter systems. The intentions behind the review of these topics are multifold. One is to introduce the aspects of Hawking-Unruh effect to the unfamiliar readers, especially from a condensed matter background and to highlight their fundamental importance. Second, the IHO is important in its own right and the discussion of such an simple quantum mechanical model has been absent from most texts. Apart from introducing the readers to these topics, the review serves to give a broader picture of the deeper structures spanning different sub-topics of physics.

\section{Inverted Harmonic Oscillator (IHO) physics in the lowest Landau level}\label{Dynamics}
 
 
We commence our exposition with a discussion on how the IHO can emerge in the context of the quantum Hall system. As is well known, the quantum Hall effect--characterized by chiral edge states and a quantized Hall conductance--can be observed in a two dimensional electron gas when subjected to a perpendicular magnetic field. At the single-particle level, the applied magnetic field leads to a discrete spectrum of evenly spaced degenerate Landau levels. Of relevance here, projecting onto the lowest Landau level (LLL) leads to non-commutativity of guiding center coordinates. 
 
 As we shall see, the IHO emerges naturally when this non-commuting nature of the LLL is combined with the presence of a saddle potential. The saddle potential is in fact ubiquitous in the quantum Hall setting \cite{Fertig87,Smitha10,Varsha19}. In the presence of disorder, it mediates quantum tunneling between equipotential trajectories\cite{Chalker}. In many experimental situations, for instance involving shot noise and anyon-interferometry \cite{Nakamura20,Bartolomei20,Halperin11,Rosenow16,Law06,Kim05,Chamon97}, the IHO is crucial to the description quantum point contacts employed for tunneling between edge states through the bulk. The saddle potential is also associated with area-preserving deformations, which are directly related to the Hall viscosity and highlight the quantum geometry associated with the system.
  
 
  
 
 
 
In order to study saddle potentials in quantum Hall systems, we begin with the Hamiltonian of a charged, free particle in a magnetic field in 2D. 
 given by
     
\beq\label{Hamiltonian}
    H= \frac{\hbar^2}{2m} \bigg( \frac{1}{i} \vec\nabla+ e\vec{A} \bigg)^2 
 \eeq
  In terms of gauge-independent ladder operators 
  \begin{align}
  b &=\frac{1}{\sqrt{2\hbar eB}}(\pi_x + i \pi_y), \\
   [b,b^\dagger]&=1 \quad \pi_i=(1/i)\partial_i+eA_i
  \end{align}

  the Hamiltonian takes the form
  \begin{align}
  H&=\hbar\omega_c (b^{\dagger}b+\frac{1}{2}),
  \end{align}
 with the cyclotron frequency
 \begin{equation}
     \omega_c=\frac{eB}{m}
 \end{equation}
  
 The guiding center coordinates, describing the centers of the electron cyclotron orbits, can be written as
   \beq
R_i =x_i +\frac{1}{eB}\epsilon_{ij} \pi_j.
\eeq
In the LLL, the two components of the guiding center operators do not commute, but instead satisfy
 \beq
 [R_i,R_j]=-i \ell_B^2 \epsilon_{ij},
 \eeq

 where we have introduced the magnetic length
 \begin{equation}
     \ell_B=\sqrt{\frac{\hbar}{eB}}.
 \end{equation}
On the other hand, the guiding center coordinates commute with the kinetic momenta:  $[R_i,\pi_j]=0$. The guiding center coordinates can be employed to construct the following ladder operators:
  \begin{align}
  a&= -i \frac{1}{\sqrt{2\ell_B^2}}(R_x -i R_y)\\
  [a,a^\dagger]&=1 \textnormal{, }
  [a,b]=[a^\dagger,b]=0
  \end{align}
Any applied potential can be represented in terms of the $a$ and $b$ ladder operators. In particular, we turn our attention to the saddle potential.
 
 \subsection{Saddle potential: Gauge invariant derivation}\label{sec:saddle}
    It was shown by Fertig and Halperin \cite{Fertig87} that the Hamiltonian for electrons in two dimensions in the presence of a high magnetic field and a saddle potential splits into two commuting parts. One part corresponds to a harmonic oscillator and the other to an inverted harmonic oscillator. The tunneling between the semi-classical orbits is completely determined by the tunneling across the inverted harmonic potential.
    Here we provide a gauge invariant derivation of Fertig and Halperin's result.
     
     The Hamiltonian for the quantum Hall system in a saddle potential is given by
     
\beq
    H= \frac{\hbar^2}{2m} \bigg( \frac{1}{i} \nabla+ \vec{A} \bigg)^2 + \lambda (x^2 -y^2).
 \eeq
 In the terms of the ladder operators defined above, the Hamiltonian reads
\beq
H=\frac{\hbar\omega_c}{2	} b^{\dagger}b +\lambda\ell_B^2(a^2 +(a^{\dagger})^2+b^2 +(b^{\dagger})^2-a^{\dagger}b -b^{\dagger}a)
\eeq
 The $a$ and $b$ operators are coupled here and can be decoupled via a rotation of basis that preserves the underlying commutation rules
 
\begin{equation}
\left(\begin{array}{c}
  a \\
  b \end{array}\right)= \left(\begin{array}{cc}
 e^{i\phi/2} \cos(\theta)&
  \sin(\theta)\\
   -\sin(\theta)  & e^{-i\phi/2}\cos(\theta)\end{array}\right)   \left(\begin{array}{c}
  c_1\\
  c_2\end{array}\right)
\end{equation} 

The choice of $\phi=0$ and $\tan(2\theta)=-4\lambda\ell_B^2/(\hbar \omega_c)$ removes the cross terms of the type $c_1^{\dagger}c_2$, $c_1c_2$., and the Hamiltonian reduces to
 \begin{align}
 H&= \frac{\hbar\omega_c}{2}\Omega c_1^{\dagger}c_1+ \lambda\ell_B^2(c_1^2 +(c_1^{\dagger})^2) \nonumber \\
 &- \frac{\hbar\omega_c}{2}(\Omega - 1)c_2^{\dagger}c_2 +\lambda\ell_B^2(c_2^2 +(c_2^{\dagger})^2)
 \end{align}
Here, $\Omega= \frac{\tan^2\theta}{\tan^2 \theta -1}$.
 We can perform a Bogoliubov transformation to diagonalize a part of the Hamiltonian with the choice of $\tanh(2 \theta_1) =-\frac{\hbar \omega_c\Omega}{ 4\lambda\ell_B^2}$ and $\tanh(2 \theta_2)= \frac{4\lambda\ell_B^2}{\hbar\omega_c(\Omega -1)}$.
\begin{equation}
\left(\begin{array}{c}
  c_i \\
  c_i^{\dagger} \end{array}\right)= \left(\begin{array}{cc}
 \cosh \theta_i &
  \sinh(\theta_i)\\
   \sinh(\theta_i)  & \cosh(\theta_i)\end{array}\right)   \left(\begin{array}{c}
  \gamma_i\\
  \gamma_i ^{\dagger}\end{array}\right)
\end{equation}
The Hamiltonian reduces to the form $H=H_1 +H_2$, where
\begin{equation}
H_1 =E_1 (\gamma_1^2 + {\gamma_1^{\dagger}}^2  )+\textnormal{constant},
\end{equation}

\begin{equation}
H_2 =E_2 (\gamma_2^{\dagger}\gamma_2 +1/2)+\textnormal{constant}
\end{equation}

where $E_1=\frac{\lambda \ell_B^2}{\cosh{2\theta_1}}$ and $E_2=\frac{-4\lambda \ell_B^2}{\sinh{2\theta_2}}$. We see that $H_1$ corresponds to a squeezing operator whereas $H_2$ corresponds to the harmonic oscillator. Making another transformation $X=(\gamma_1^{\dagger}-\gamma_1)/(\sqrt{2}i)$, $P=(\gamma_1+\gamma_1^{\dagger})/\sqrt{2}$ and $x=(\gamma_2^{\dagger}+ \gamma_2)/\sqrt{2}$, $p =(\gamma_2- \gamma_2^{\dagger})/(\sqrt{2}i)$, we obtain
 \begin{equation}
 H=E_1 (P^2 -X^2) + E_2/2 (p^2 +x^2).
 \end{equation}
 Thus the Hamiltonian for the quantum Hall system in a saddle potential is a sum of an inverted oscillator and a harmonic oscillator, a result similar to that obtained in \onlinecite{Fertig87}, but derived here in a manifestly gauge invariant form. In the limit $B \rightarrow \infty$, the system is restricted to one of the Harmonic oscillator levels, equivalent to projecting onto the lowest Landau level. In terms of guiding center co-ordinates, the Hamiltonian in the lowest Landau level is the inverted harmonic oscillator.

The saddle potential serves well to model the bulk potential energy profile in point contacts mesoscopic quantum Hall devices, created in pinched geometries that bring edge states close together\cite{Fertig87,Buttiker90}.  The conductance in such a point contact geometry is expressed in terms of transmission probability of single particle states (and more generally, quasiparticles) under the influence of a saddle potential \cite{Fertig87}. Here we have shown that this problem is reduced to transmission across the IHO barrier. The transmission co-efficient $|t|^2$ can be exactly computed in this set-up\cite{Buttiker90,Fertig87,Hegde19}, yielding the well-known formula
\begin{equation}
 |t|^2 =\frac{1}{1+e^{-2\pi \epsilon}},
\end{equation}
where $\epsilon \sim E/\lambda$.
  The transmission coefficient is completely determined by the physics of the inverted harmonic oscillator, and takes a form reminiscent of a thermal (Fermi-Dirac) distribution. In subsequent sections, we will derive this form using a scattering formalism as well as relate it to the thermal nature of quantum states near an event horizon. We now  examine the algebraic structure of the  class of potentials that yield an IHO when projected to the lowest Landau level.

 \subsection{Electrostatic potentials and strain generators in LLL}
 
  
 The saddle potential is prevalent in two quantum Hall contexts:   i) Generators of strain that preserve flux play a key role attributed to geometric deformations, be it as a tool for deriving the form of response function or as can be elicited by the application of stress in recent experiments.ii) As discussed in the previous subsection, for potential landscapes that are shallow compared to the Landau level spacing and on large scales compared to the magnetic length, local variations can generally be captured by quadratic potentials. Here we study both cases with regards to their algebraic structure and projection to the lowest Landau level. 
  
\subsubsection{Strain generators}
Geometric deformations in a quantum Hall system amount to uniform area preserving deformations of a two dimensional system in a magnetic field. In order to obtain the associated strain generators, consider the transformations on $(x_i,\pi_i)$ obtained by generators $J_{ij}$: $S=e^{-i\Lambda_{ij}J_{ij}}$ , $\Lambda =e^\lambda$ such that $S x_i S^{-1}=\lambda_{ji}x_j , S \pi_i S^{-1} = \Lambda^{-1}_ {ij} \pi_j $. 	This gives us the algebraic relations:
 \bea
 i[J_{ij}, \pi_k]=\delta_{ik} \pi_j  \\
 i[J_{ij}, x_k]= -\delta_{jk} x_i  \\
 i[J_{ij}, J_{kl}] =\delta_{il} J_{kj} -\delta _{jk}J_{il}
 \eea
  The last condition defines the algebra of these generators, which is the $\mathfrak{sl}(2,\mathbb{R})$ Lie algebra.
 The strain generators can be written in terms of the $(x_i,\pi_i)$ from the above conditions \cite{ReadVisc,Bradlyn2012}:
  \beq
  J_{ij}= -\frac{1}{2\hbar} \{x_i,\pi_j\}+\frac{1}{4\hbar} \{x_i,\pi_j \}\delta_{ij}+\frac{1}{2\ell_B^2} \epsilon_{ik}x_j x_k
  \eeq
  The first two terms generate shear in the absence of a magnetic field. This can be seen from the fact that $\pi_i$ are the generators of  `kinetic translations'. The last term appears in the presence of a magnetic field as gauge transformations and also  compensates for the non-commutativity of kinetic momenta.

  The rotation generator is given by:
  \beq\label{eq:j1}
  L/\hbar=-\frac{1}{2}\epsilon^{ij} J_{ij} =-\frac{1}{4} \bigg(  \frac{1}{\ell_B^2}|\vec{R}|^2 +\frac{\ell_B^2}{\hbar^2}|\vec{\pi}|^2  \bigg)
  \eeq
  Two shear generators therefore take the form
  \beq\label{eq:j2}
  J_a = \frac{1}{2} \sigma^z_{ij}J_{ij} =\frac{\ell_B^2}{4\hbar^2} \{\pi_x,\pi_y \}+\frac{1}{4\ell_B^2} \{ R_x,R_y \}
  \eeq
  \beq\label{eq:j3}
  J_b =\frac{1}{2}\sigma^x_{ij}J_{ij} =\frac{1}{4\ell_B^2} (R_y^2 -R_x^2) + \frac{\ell_B^2}{4\hbar^2} (\pi_y^2 - \pi_x^2)
  \eeq
  Therefore, one can see that the guiding center and the kinetic parts of the generators decouple. On restricting to the lowest Landau level, one is retained with guiding center part only. We'll see that the three generators correspond to harmonic oscillator, inverted harmonic oscillator and dilatation generator Hamiltonians in the lowest Landau level.

 \subsubsection{Quadratic potentials--}

We can perform a similar algebraic analysis of a generic quadratic potential in a quantum Hall system. By decomposing the tensor $x_ix_j$, just as we did for $J_{ij}$, we can enumerate the three linearly independent quadratic potentials in two dimensions. The are
 \bea
V_1 =\lambda_1 (x^2+y^2),  \quad 
V_2= \lambda_2(xy), \quad
V_3=\lambda_3 (x^2 -y^2)
 \eea
 
One can restrict to the LLL to study the electron/quasiparticle dynamics in the presence of high magnetic fields. To this end, consider a Hamiltonian of the form 



    \beq
    H=H_0 +V_i =  \frac{1}{2m} \pi_i^2 +V_i =\hbar\omega_c \bigg( b^{\dagger}b +\frac{1}{2} \bigg)+V_i
    \eeq
  
We now introduce the lowest Landau level projection operator $P_{LLL}$ that satisfies the following relations with Landau level lowering/raising operators and the angular momentum operators:$[b,P_{LLL}]=-P_{LLL}b,  \quad [a,P_{LLL}]=0$
   For a normal ordered function of $a,b$ operators, $f(b,b^\dagger,a,a^{\dagger})= \sum_{n,m} g_{nm}(a,a^\dagger)(b^\dagger)^n b^m$, the LLL projection is then given by $P_{LLL}f(b,b^{\dagger}, a ,a^{\dagger}) =g_{00}(a,a^{\dagger})$
   The $a$ operators are given only in terms of the guiding center co-ordinates  :  $a= -i 1/\sqrt{2\ell_B}(R_x -i R_y)$ and the $b$ operators are similarly given in terms of the kinetic momenta $\pi_i$. The projection to LLL leaves us with expressions only involving $R_x,R_y$. one can see that LLL projections of the potentials $V_i$ are given by
   \begin{align}
        P_{LLL}V_1 P_{LLL} &= \lambda(R_x^2 +R_y^2)+\mathcal{O}(1)\\
   P_{LLL}V_2 P_{LLL}&= \frac{\lambda}{2} (R_x R_y + R_y R_y ) + \mathcal{O}(1) \\
      P_{LLL}V_3 P_{LLL} &= \lambda(R_x^2 -R_y^2) + \mathcal{O}(1)
      \label{Eq.LLLpots}
   \end{align}

 Furthermore, we can also project the strain generators to the lowest Landau level to find $P_{LLL}J_{ij}P_{LLL}= \frac{1}{4\ell_B^2}\epsilon^{jk} \{ R_i, R_k\} + \mathcal{O}(1)$. From this, we see that
 \begin{align}
 P_{LLL}V_1P_{LLL}&=-4\lambda\ell_B^2/\hbar P_{LLL}LP_{LLL} \\
  P_{LLL}V_2P_{LLL}&=2\lambda\ell_B^2 P_{LLL}J_aP_{LLL} \\
   P_{LLL}V_3P_{LLL}&=-4\lambda\ell_B^2 P_{LLL}J_bP_{LLL} 
   \end{align}

    Thus, we see that the strain generators and the electrostatic potentials lead to the identical quadratic Hamiltonians when projected to the LLL.  This is due to the fact that both the strain generators and the bilinears $R_iR_j$ are generators of the algebra $\mathfrak{sl}(2,\mathbb{R})$. From the above, it can be seen that on projection to the LLL, the kinetic terms drop out and the potentials $V_i$ act as the Hamiltonians acting on the LLL states \cite{girvinjach}.

    
    We are therefore left with three simple quadratic potentials that generate the Hamiltonian dynamics in the LLL. As one of the set, the IHO arises naturally in the quantum Hall context. In what follows, we perform a comprehensive analysis of the IHO and its scattering properties, as well as parallels between the LLL description and rotations and Lorentz transformations in relativistic Minkowski descriptions.

  \begin{figure}
\centering
\includegraphics[width=0.5\textwidth]{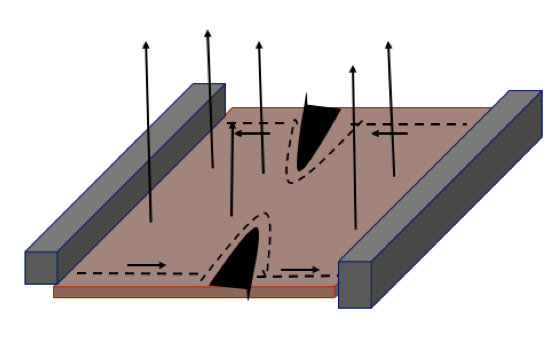}
\caption{A quantum Hall system comprises of a two dimensional electron gas in the presence of a strong magnetic field. The states at the edges have a chiral nature and are unidirectional.  Point contacts are applied as probes for conductance measurements and are modeled with a saddle potential $V(x,y)=\lambda(x^2-y^2)$.}
 \label{fig:Saddle}
\end{figure}

 \begin{figure}
\includegraphics[width=0.4\textwidth]{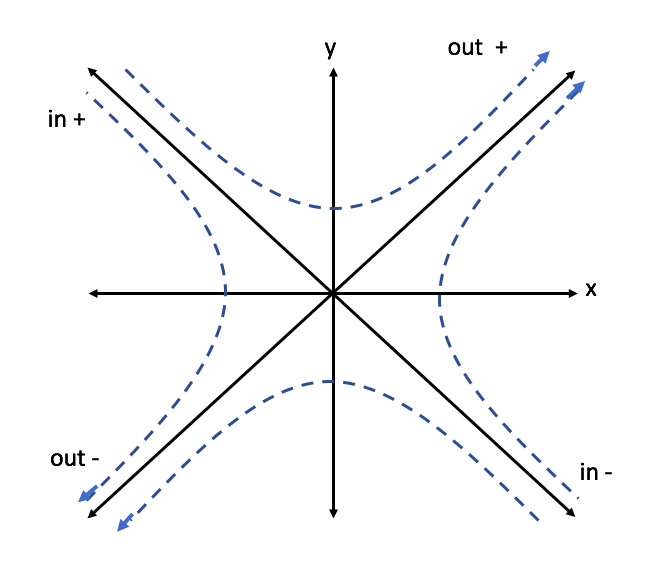}
\includegraphics[width=0.4\textwidth]{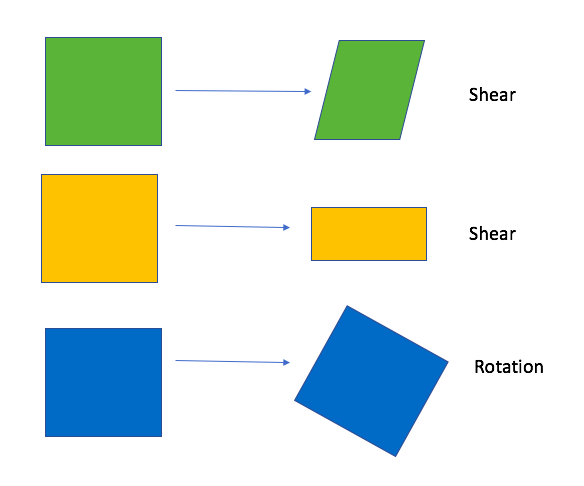}

\caption{The top figure shows the semiclassical trajectories for the IHO in phase space/non-commutative plane of the LLL. $u^\pm$ indicate the `light-cone' basis obtained through a canonical transformation from the $(X,P)$ basis. The state in the $u^\pm$ basis are purely incoming and outgoing states. The dotted curves are the hyperbolic trajectories of constant energy. The bottom figure shows the 3 different area preserving transformations applied to the quantum Hall system in the lowest Landau level. The shear transformation  manifests as the IHO on projection to the LLL and the single particle states in the quantum Hall system follow the above semi-classical trajectories. }
 \label{fig:Shear}
\end{figure}

\section{Quantum mechanics of the  Inverted Harmonic oscillator}\label{IHO}
Having discussed the importance of the inverted Harmonic oscillator (IHO) in the lowest Landau level context, here we present a detailed review of the quantum mechanics of the IHO, highlighting the features that will be of importance in further sections. We start with a quick survey of works on the IHO, present it in different forms and physical manifestations, then study the properties of eigenmodes, the scattering matrix and finally the decaying states of the model.

Various aspects of the IHO have been studied in differing degrees of depth and in multitudinous contexts. Yet this simple, exactly solvable model has fallen out of fashion in textbook discussions, even though it could very well serve as an archetypal counterpart of the simple harmonic oscillator(SHO). The SHO highlights many characteristic features of quantum mechanics such as the existence of a ground state, eigenvalue quantization and other aspects of bound system in their simplest form. Similarly, IHO brings forth fundamental aspects of quantum dynamics such as tunneling, decay and other aspects of an open, scattering system.   The ubiquitous nature of the IHO could also be gleaned from considering it as a `saddle-point' approximation to maxima in potential landscapes, just the way the SHO appears as a 'saddle-point' approximation for potential minima. Moreover, as we focus on in the following discussion, the IHO manifests many non-trivial quantum mechanical aspects in terms of its scattering amplitudes, time-evolution, and the existence of time-decaying resonant states, to name a few.


The quantum mechanics of the IHO and its scattering properties have been studied in various contexts and in depths since the 1930s up to this day. \cite{Kemble37,Landau, Barton86,Chruscinski03,Chruscinski04,Yuce,Shimbori00,Shimbori1,Shimbori,Berry,Sierra,Bhaduri,Bhaduri97,Hashimoto20,Bhattacharyya19,Bhattacharyya20,Ali20,tian20,Betzios20, Dalui20,Dalui19,Dalui192,Hegde19}. We shall give a thorough survey of the various fundamental contexts in which this model appears, ranging from cosmological inflation, black holes, quantum optics to string theory in Sec.\ref{survey}.

The presence of IHO physics in the context of Hawking-Unruh effect was already pointed out in some works in string theory \cite{Cremonini,Friess,Maldacena05}. More recently, the black hole S-matrix was shown to be related to that of the IHO. \cite{Betzios16} 
The realisation of the IHO in quantum Hall systems and its relation to the Lorentz boost and the Rindler Hamiltonian was pointed by the authors recently\cite{Hegde19}. Since then there have been several recent works which have highlighted the appearance of IHO in the context of horizon thermality and chaos \cite{Hashimoto20,Bhattacharyya19,Bhattacharyya20,Ali20,tian20,Betzios20, Dalui20,Dalui19,Dalui192}.  



From this brief survey of topics, the importance of the IHO is apparent in its applicability to diverse physical phenomena, acting as a prototypical model in bringing out key quantum mechanical aspects. Now we undertake a brief review of the quantum mechanics of the IHO specifically highlighting the aspects relevant for further discussions.

\subsection{IHO in different forms: A scattering potential, a dilatation generator and  a squeeze generator}\label{IHOforms}


In the case of ubiquitous simple harmonic oscillator studies, representing the underlying Hamiltonian in different bases goes far in providing insights for different aspects and settings, from wavefunctions in a trapping potential to operators associated with coherent states. Likewise, the Hamiltonian operator for the IHO has different representations each related to one another through a canonical transformation. We present here several useful bases: the position basis $(x,p)$, the 'light-cone' basis $(u_+,u_-)$ and the operator basis $(a,a^\dagger)$. As we now show, analogous to the SHO, these representations of the states may be interpreted as a scattering potential, a dilation generator or a squeeze generator, depending on the basis and manifold of choice. These different perspectives and their common root demonstrate the interconnected nature of seemingly different phenomena in diverse sub-fields. 

The Hamiltonian for an inverted oscillator potential, which forms the basis of scattering,  can be written as 
  \beq
 H= \frac{1}{2m}p^2-\frac{m\lambda^2}{2} x^2 \label{ldef}
 \eeq

 Here $p$ and $x$ are the usual momentum and position operators acting on the Hilbert space. Here, the curvature of the potential is given by $\lambda$ analogous to the energy scale $\omega$ in the SHO. In the context of IHO, it also sets the time scale associated with the scattering process against the barrier.

 
 In the position basis, the Hamiltonian results in the following Schr\"odinger equation for the energy eigenvalue $E$:
 \beq
 H \psi(x)=  \frac{1}{2m} \bigg(- \hbar^2\frac{\partial^2}{\partial x^2} - m^2\lambda^2 x^2 \bigg)\psi(x)=E\psi(x)\label{ihoschrodinger}
 \eeq
 This equation describes scattering off a parabolic potential barrier, in contrast to a trapping potential in the SHO. It is known to have a  solution described by the Weber equation which can be expressed in terms of special functions\cite{Gradshteyn}. Being a scattering problem, the natural interpretation of these solutions are as scattering modes, characterized by a scattering matrix (S-matrix), as shown in Sec. \ref{Stheory}.

 A more convenient basis entails the `light-cone basis'.
 A canonical transformation from the position basis to the `light-cone' basis $u^{\pm}$ given by:
\begin{equation}
u^{\pm} =\frac{p \pm m\lambda x}{\sqrt{2m\lambda}}
\end{equation}
The canonical commutation relation is preserved under the transformation : $[u^+, u^-]=i\hbar$. In this basis the Hamiltonian takes the form
\begin{eqnarray}
H&=\frac{\hbar\lambda}{2} (u^+ u^- +u^- u^+) \label{eq:lightconeH}\\
&=  \pm i \lambda\hbar \bigg( u^{\pm} \partial_{u^{\pm}} + \frac{1}{2}\bigg)
\end{eqnarray}
The associated basis states $\ket{u^{\pm}}$ in this basis correspond to the incoming (outgoing) states towards(away from) the barrier. 
 We recognize this Hamiltonian as the generator of dilatations of the light-cone coordinates. The eigenfunctions of the dilatation generator are given by power functions of the form $(u^\pm)^n$. Equation.~(\ref{eq:lightconeH}) generates a scale transformation on the functions it acts on \cite{FrancescoCFT}. 
 Equation.~\ref{eq:lightconeH} is also known as the `Berry-Keating' Hamiltonian studied in relation to quantum chaos \cite{Berry,Sierra}. We will be extensively using `light-cone' basis throughout in the following sections.We will also make a direct comparison of the eigenstates of IHO in this basis to the quantum mechanical modes near an event horizon(Rindler modes) in Section~\ref{RindHam}. 

The IHO may also be viewed as a generator of squeezing, which becomes manifest upon considering its phase space dynamics. The squeeze operator is formally written in terms of the creation and annihilation operators $a, a^{\dagger}$. These operators are defined as $a=\sqrt{\frac{\lambda}{2\hbar}}(u^++i\frac{u^-}{\lambda})$,$a^{\dagger}=\sqrt{\frac{\lambda}{2\hbar}}(u^+-i\frac{u^-}{\lambda})$ and obey $[a,a^{\dagger}]=1$,$[a,a]=0$. The IHO Hamiltonian is then given by
 \beq
 H=i((a^{\dagger})^2 -a^2)
 \eeq
 The `vacuum' state is defined as $a\ket{0}=0$. In the context of the SHO, the vacuum state is the zero-point energy eigenstate of the Hamiltonian, which under the action of the creation operator gives rise to higher quantized energy eigenstates. Since we are dealing with a scattering problem with an unbounded potential, there is no such state. Rather, when the vacuum state is evolved in time via the IHO Hamiltonian, it becomes a ``squeezed state''\cite{Teich,Walls1983}. To obtain some intuition for this, consider the action of a squeezing operator on a SHO coherent state, which satisfies minimum uncertainty and has equal spread in (renormalized) position  and momentum. The action of the squeeze operator results in a state having reduced spread in one phase space direction, and elongated spread in the perpendicular direction, such that the product of the widths in position and momentum (the area in the phase space) remains constant.
 
 Formally, the most general squeeze operator $S(z)=\exp(\frac{1}{2}(z(a^{\dagger})^2 -z^{*} a^2))$  is parametrized by a complex number $z=re^{i\theta}$, and transforms the ladder operators as
 \bea
 b&\equiv S^{\dagger}aS \equiv (\cosh{r}) a+ (e^{i\theta} \sinh{r})a^{\dagger} \\
  b^{\dagger}&\equiv S^{\dagger}a^{\dagger}S = (\cosh{r}) a^{\dagger}+ (e^{-i\theta} \sinh{r})a. \label{squeeze}
 \eea
 The squeeze operator mixes the creation and annihilation operators and leads to a Bogoliubov transformation that still preserves the canonical commutation relations $[b,b^{\dagger}]=1$, $[b,b]=0$ but changes the `particle number' content of the vacuum state. That is, the action of the squeeze operator on the vacuum becomes
 \begin{align} \ket{z}&=e^{\frac{1}{2}(z \hat a^2-z^*\hat {a^\dagger}^2)}\ket{0} \\
&=\frac{1}{\sqrt{\cosh r}}\sum_{n=0}^\infty (-e^{i\phi}\tanh r)^n \frac{\sqrt{(2n)!}}{2^n n!}\ket{2n}
\end{align}

It may be verified that the average occupation number $<n>=<a^\dagger a>$ of the squeezed vacuum is no longer zero, and instead takes the value $<n>=\sinh^2 r$. This may be expressed as
 \beq
 \bra{z}n\ket{z}=\bra{z}a^\dagger a\ket{z}=\bra{0}b^{\dagger}b \ket{0}= \frac{1}{|\tanh{r}|^{-2}-1}
 \eeq
 As we will see in Section~\ref{RindHam}, this relation manifests as Hawking radiation or the Unruh effect in the context of spacetime horizons.
 
 Finally, we note that the squeezing and dilation generators are intimately related to the algebra of Lorentz transformations. Consider the phase space, of say, a photon or a one dimensional quantum harmonic oscillator. The geometry of this space is defined by the commutation relation $[x,p]=i\hbar$.  The phase space is thus left invariant under symmetry transformations generated by the symplectic Lie algebra $\mathfrak{sp}(2,\mathbb{R})$.    One of the symplectic transformations are given by
\begin{align}
    x &\rightarrow x \cosh \beta + p \sinh \beta\\
    p &\rightarrow x \sinh \beta +p \cosh \beta.
\end{align}
We note that this transformation is identical to a Bogoliubov transformation between  quantum mechanical operators as well as to a "Lorentz boost" on a spacetime manifold. This is an effect of the local isomorphism between the symplectic group and the indefinite orthogonal group in 2+1 dimensions. That is, $\mathfrak{sp}(2,\mathbb{R})\sim\mathfrak{so}(2,1)$. This isomorphism is crucial in connecting the IHO to aspects of black hole thermality \cite{Bishop86} in later sections.  In "lightcone" coordinates $x_\pm\sim x\pm p$, the same transform is given by $x_\pm \rightarrow x_\pm e^{\pm\beta}$. We may identify the generator for this dilation transform as  
   \beq e^{i\beta(-iK_1)}x_\pm=e^{\pm\beta}x_\pm, \label{scale}\eeq 
 where
 \beq -iK_1 \sim x_+x_-+x_-x_+ \sim p^2-x^2. \eeq 
 These "boost" generators in phase space execute hyperbolic trajectories, and take the same form as the Hamiltonian for a one dimensional inverted harmonic oscillator. Note that the other two generators of this algebra are another boost $K_2\sim xp+px$ and a rotation $K_3\sim x^2+p^2$. Henceforth, the two boost generators will be equivalently referred to as IHO Hamiltonians. 

 Having seen the different avatars of the IHO, we proceed to examine the properties of the Hamiltonian.

 \subsection{Properties of the evolution operator}
 
Here, we review certain basic features of the IHO Hamiltonian, such as time-evolution, self-adjointness and quantum mechanical PCT (Parity, Charge conjugation, Time reversal)  symmetries, highlighting the uniqueness of the IHO in these regards.

 {\it Time evolution operator -- } The IHO has the characteristic feature that time-evolution of an appropriate wavepacket distribution leads to temporal decay characterized by quantized decay rates. First considering evolution within the regular Hilbert space, we employ the `light-cone' basis : $H= i \lambda(u \partial_u+\frac{1}{2})$ ({where we have suppressed the $\pm$ label of the previous subsection and set $\hbar=m=1$ for convenience of notation}). 
 We define the operator:
 \beq
 U=e^{-iH t}= e^{\lambda t/2} e^{\lambda t u \partial_u}.
 \eeq
 This operator acts on functions $\psi$ belonging to the Hilbert space as a dilatation operator, leading to an isometry on the Hilbert space \cite{Chruscinski03,Marcucci16}:
 \beq
 U \psi(u) = e^{\lambda t/2} \psi(e^{\lambda t}u).
 \label{Eq:TimeEvol}
 \eeq
  The non-unitary appearance of this transformation is not a concern as the stronger requirement of self-adjointness is satisfied as shown below. Thus, the IHO generates a dynamical scaling action on the wavefunctions, termed as `modularity' \cite{Padmanabhan19}. This behaviour is identical to the dilatation of the time-coordinate near an event hoirzon of a black hole and the resulting red-shift of quantum mechanical modes \citep{Padmanabhan19}.

  The self-adjointness of the Hamiltonian is a necessary condition for the unitary evolution in quantum mechanics.  Typically, in finite dimensional cases, self-adjointness is equivalent to the Hermitian ($H^\dag=H$) or symmetric property of operators. One needs to be careful in infinite dimensional cases and for unbounded operators such as the IHO Hamiltonian. The momentum operator in the position representation is the simplest example where one is not guaranteed the self-adjoint property as a consequence of the Hermitian property. Here we will prove the self-adjoint property of the IHO with respect to the Hilbert space of square-integrable functions $L^2$. From Eq.\ref{Eq:TimeEvol}, one can show that for $\psi , \phi \in L^2$ :
 \bea
 \non  \bra{U \psi}| U \phi \rangle &=& \int_{- \infty}^{\infty}du \overline{U \psi(u)} U \phi(u) = \int_{- \infty}^{\infty}du e^{t}\overline{ \psi(e^{t}u)} \phi(e^t u) \\ 
& =&\int_{- \infty}^{\infty} \bar{\psi(y)} \phi(y)dy=\bra{\psi} \phi \rangle,
 \eea
  where in the last line we have made the substitution $y=e^tu$. This ensures unitarity of $U$, and the self-adjointness of $H$ follow from Stone's theorem (which states a correspondence between self-adjoint operators on Hilbert space and one parameter family of unitaries \cite{Chruscinski03}). 
 
{\it PCT operations--} The IHO has noteworthy time-reversal properties for a simple quantum mechanical Hamiltonian. As shown by Wigner, the T operator can be realised either as a unitary or an anti-unitary operator.  The IHO is an unbounded operator and with a unitary T, satisfies the relation \citep{Chruscinski03}
\beq
TH+HT=0
\eeq
 Therefore, 
  \beq
  H\ket{\psi_E}=E \ket{\psi_E}, \quad HT\ket{\psi_E} =-ET \ket{\psi_E}
  \eeq
   To see this consider $\ket{\psi(t)}= U(t)\ket{\psi}$, where $U(t)$ is the time-evolution operator. The evolution of the time-reversed state is given by : $T( U(t)\ket{\psi})=U(-t)(T\ket{\psi})$.  In the case of bounded Hamiltonians, one chooses T to be anti-unitary to exclude negative energy eigenvalues. But this is not the case for the IHO and we can choose a unitary time-reversal operation.
  As a consequence, one obtains two families of states in the energy spectrum $H \psi_{\pm}=\pm E \psi_{\pm}$, where the $\psi_\pm$ are related to by the T operator. That is, $T\psi_E=\psi_{-E}$. That is, IHO has both negative and positive energy spectra.
  
  
  Further, defining the parity operator as : $PxP^{-1}=-x$ and $PpP^{-1}=-p$ , one can see that the IHO Hamiltonian is P-symmetric:
\beq
PHP^{-1}= H.
\eeq

 Now let us consider the complex conjugation operator C defined as : $C \psi =\bar{\psi}$. The IHO Hamiltonian is both CT and PCT invariant \cite{Chruscinski03}:
  \beq
  [H,CT]=[H,PCT]=0
  \eeq


  \subsection{Eigenmodes: In- and out-going states}\label{sec:parabolic-cylinder}
  The PCT properties of the IHO allow for some noteworthy properties in the spectrum of the Hamiltonian. As the Hamiltonian is unbounded, the spectrum of real energy eigenvalues is continuous and ranges from $-\infty$ to $\infty$. The parity invariance leads to a doubly degenerate spectrum. This is associated with the states on the two sides of the barrier.

   The $u^+$ basis describe the ingoing states and $u^-$ the outgoing states and these two basis are related by\citep{Betzios16, Maldacena05}:
\begin{equation}
\bra{u^+}  u^- \rangle = \frac{1}{\sqrt{2 \pi}} e^{iu^+ u^-}
\end{equation}

 The time-dependent Schrodinger equation, in either basis, is  of the form 
 \begin{equation}
 i \partial_t \psi_{\pm}(u^{\pm},t)  =\epsilon \psi_{\pm}= \mp i \lambda(u^{\pm}\partial_{u^{\pm}}+1/2)\psi_{\pm}(u^{\pm},t).
 \label{epsilonE}
 \end{equation}
{In the following, we will use the scaled energy $E=\epsilon/\lambda$ and set $\hbar=m=1$ for convenience of notation.}
 
As shown in the Fig.\ref{fig:Shear}, there are two sets of energy eigenstates corresponding to regions I and II. The state $\ket{E, \pm}$ correspond to the states in the two regions respectively. These are written in the in-going and out-going bases. In terms of in-going bases we have, 
\begin{equation}
\bra{u^+} E,+ \rangle _{in} =\frac{1}{\sqrt{2 \pi}}(u^+)^{iE-1/2} \Theta(u^+)
\label{Eq:Mode1}
\end{equation}
\begin{equation}
\bra{u^+} E,-\rangle _{in} =\frac{1}{\sqrt{2 \pi}} (-u^+)^{iE-1/2} \Theta(-u^+)\label{plus}
\end{equation}
where $\Theta(u^{+})$ is the Theta-step function. In terms of outgoing bases, we have
\begin{equation}
\bra{u^-} E,+ \rangle _{out} =\frac{1}{\sqrt{2 \pi}}(u^-)^{-iE-1/2} \Theta(u^-)\label{minus}
\end{equation}
\begin{equation}
\bra{u^-} E,- \rangle _{out} =\frac{1}{\sqrt{2 \pi}}(-u^-)^{-iE-1/2} \Theta(-u^-)
\label{Eq:Mode4}
\end{equation} 
  These equations correspond to the `steady-state' scattering states. In the position basis, we can write these as 
   \begin{align}
   \chi_E(x)&=\langle x\ket{E,+} = \int du^+ \langle x \ket{u^+}\langle u^+ \ket{E,+}\\
   &= N_0 e^{\pi \frac{E}{4}} \Gamma(\frac{1}{2}-iE) D_{iE-\frac{1}{2}}(x)
   \end{align}
 where $D_\nu(x)$ are parabolic cylinder functions \cite{Zwillinger}. Normally, to solve for the scattering matrix of a barrier potential, one would assume incident plane waves at infinity, where the barrier potential has no support and hence, only the kinetic energy term remains in the Hamiltonian. But the IHO has support throughout the real line, and hence, one cannot consider plane waves as asymptotic solutions of the scattering states. Instead, the `plane-wave-like' asymptotic states go as $e^{\pm ix^2}$\cite{Barton86,Chruscinski03,Chruscinski04}.

 Thus, any wavepacket that has to be constructed to scatter off the IHO must be expressed in terms of these asymptotics, as we shall do in Sec.\ref{wpscat}. Other than scattering states, the IHO also allows for resonant states that have complex eigenvalues. The nature of all these states may be gleaned from the scattering matrix, which we now compute.
 
\subsection{S-matrix for the IHO : Mellin Transform }\label{Stheory}
Now that we have the eigensolutions, we calculate the scattering matrix for the IHO. In standard quantum mechanical scattering problems, one considers plane waves to scatter against a barrier \cite{Landau, Maldacena05}.
These plane waves are eigenmodes of the momentum operator. The spatial bounded nature of a typical scattering potential allows one to consider plane wave states at infinity.  Any other state is expanded in the basis of plane waves resulting in a Fourier transform.
  When dealing with the previously described dilatation operator form of the intrinsically unbounded IHO, the Mellin transform, which is a multiplicative version of Fourier transform, becomes important\cite{Moses72,Cycon87,Lowe16,Dhritiman12,Fitzpatrick11}. 
  The Mellin Transform is defined as 
\begin{equation}
\tilde{F}(\epsilon)= \int_0^{\infty} f(u) u^{i \epsilon-1/2}du,
\label{mellin}
\end{equation}


 From equations \ref{plus}, \ref{minus} and \ref{mellin}, it can be seen that the Mellin transform is an expansion in the basis of eigenfunctions of the IHO in light-cone coordinates . Now we shall make use of this in the derivation of the S-matrix for the IHO .

 A state going towards the IHO barrier can be expanded in terms of the eigen-solutions in the in-going basis as 
\begin{equation}
\begin{split}
&\hat{\psi}_{in}(u^+)\\=& \frac{1}{\sqrt{2 \pi}} \int_{-\infty}^{\infty} dE [(u^+)^{iE-1/2} \ket{E,+}_{in} +(-u^+)^{iE-1/2} \ket{E,-}_{in}]
\end{split}\label{instate}
\end{equation}
The mode expansion for the outgoing state is given by
\begin{equation}
\begin{split}
&\hat{\psi}_{out}(u^-)\\=& \frac{1}{\sqrt{2 \pi}} \int_{-\infty}^{\infty} dE [(u^-)^{-iE-1/2} \ket{E,+}_{out} +(-u^-)^{-iE-1/2} \ket{E,-}_{out}]
\end{split}\label{outstate}
\end{equation}

The S-matrix relates the out and in states on the two sides of the barrier as follows
\begin{equation}
\left(\begin{array}{c}
  \ket{E, +}_{out}\\
  \ket{E, -}_{out}\end{array}\right)=\hat{\mathcal{S}}\left(\begin{array}{c}
  \ket{E, +}_{in}\\
  \ket{E, -}_{in}\end{array}\right)
  \end{equation}
  
  The above-defined in and out states are then related by \cite{Maldacena05,Betzios16}
  \begin{equation}
  \hat{\psi}_{out}(u^-)=[\hat{\mathcal{S}}](u^{-})= \int_{-\infty}^{+\infty} \frac{du^+}{\sqrt{2 \pi}} e^{-i u^+ u^-} \hat{\psi}_{in}(u^+)
  \end{equation}
  
  For simplifying the above we shall make use of the Mellin transforms 
  \beq 
  \begin{split}
  \int_{-\infty}^{\infty} du^+ e^{iu^+ u^-} & |u^+|^{-iE-1/2}  = \\
  &e^{i\pi/4}e^{E \pi/2}|u^-|^{iE-1/2} \Gamma(\frac{1}{2}-iE)
  \end{split}
  \eeq
  
  \beq
  \begin{split}
  \int_{-\infty}^{\infty} du^+ e^{-iu^+ u^-} & |u^+|^{-iE-1/2} =  \\ & e^{-i\pi/4}e^{-E \pi/2}|u^-|^{iE-1/2} \Gamma(\frac{1}{2}-iE)
  \end{split}
  \eeq
  
  The S-matrix is then given by
  \begin{equation}
  \left(\begin{array}{c}
    \ket{E,+}_{out}\\
    \ket{E,-}_{out}\end{array}\right)=\hat{\mathcal{S}}\left(\begin{array}{c}
  \ket{E,+}_{in}\\\
  \ket{E,-}_{in}\end{array}\right)
  \label{Eq:OpSmatrix}
  \end{equation}
  
  \begin{equation}
  \hat{\mathcal{S}}= \frac{1}{\sqrt{2 \pi}} \Gamma\bigg( \frac{1}{2}-iE  \bigg)  \left(\begin{array}{cc}
 e^{-i\pi/4}e^{-\pi E/2}&
  e^{i\pi/4}e^{\pi E/2}\\
   e^{i\pi/4}e^{\pi E/2}& e^{-i\pi/4}e^{\pi E/2}\end{array}\right)\label{SMat}
  \end{equation}
  
  The probabilities for tunneling and reflection across the IHO barrier can then be gleaned from the above expression:
  \beq
  T=|S_{12}|^2=\frac{1}{1+e^{-2\pi \epsilon/\lambda}}, \quad R=|S_{11}|^2=\frac{e^{-2\pi \epsilon/\lambda}}{1+e^{-2\pi \epsilon/\lambda}}\label{Tval}
   \eeq
   Here, we have re-introduced $\epsilon$ which is the eigenvalue of the IHO Hamiltonian in Eq. \ref{epsilonE} to stress that the ` effective temperature' in the thermal factor is associated with the strength of the IHO potential. Note that the expression for the transmission probability $T$ here matches the well-known formula for the conductance through a point contact invoked in Sec.~\ref{sec:saddle} for the quantum Hall system. We shall later interpret these probabilities in the context of black hole thermality and further analyses of the quantum Hall context.

\subsection{Scattering states in the position basis: Parabolic cylinder functions}\label{wpscat}
As we saw in a previous subsection, the Schrodinger equation in the position basis for the IHO has the form of the Weber equation. The solutions are known to be expressed in terms of parabolic cylinder functions\cite{Abramowitz}. We can instead obtain them easily from the solutions in $u^{\pm}$ basis as
   \begin{equation}
   \chi_E(x)=\langle x\ket{E,+} = \int du^+ \langle x \ket{u^+}\langle u^+ \ket{E,+}
   \end{equation}
The canonical transformation from $(x,p)$ to $(u^+, u^- )$ operators has a corresponding representation given by \citep{Maldacena05, Chruscinski03, Chruscinski04}
\begin{equation}
\langle x \ket{u+} =\text{exp}[i(\frac{x^2}{2}-2u^+ x + {u^+}^2)]
\end{equation}  
  Therefore, there transformation now reads
  \begin{equation}
   \langle x\ket{E,+} = \int_0^{\infty} du^+ \text{exp}[i(\frac{x^2}{2}-2u^+ x + {u^+}^2)] {u^+}^{iE-1/2}
   \end{equation}
  The integral representation of the parabolic cylinder function is given by 
  \begin{equation}
  D_{iE-1/2}(x) =\frac{e^{\frac{-x^2}{4}}}{\Gamma(\frac{1}{2}-iE)}\int_0^{\infty} dt |t|^{-iE-\frac{1}{2}} \text{exp}(\frac{-t^2}{2}-xt)
  \end{equation}
  Using this, the solution in the position basis is obtained as 
  \begin{equation}
  \chi_E(x)=\langle x\ket{E,+}= N_0 e^{\pi \frac{E}{4}} \Gamma(\frac{1}{2}-iE) D_{iE-\frac{1}{2}}(x)
  \end{equation}
 One can see the uniqueness of IHO as a scattering problem from the asymptotic behaviour of the states in the position basis.  In a textbook quantum mechanical scattering problem, rather than studying the eigenstates of the Hamiltonian, one starts with a potential with finite support in a given region of space and considers scattering the `states of choice' off the potential barrier. Usually the `states of choice' are the plane waves states $e^{ikx-i\omega t}$, which are also the eigenstates of the kinetic energy operator $P^2/2m$ in the Hamiltonian. The boundedness of the potential allows one to do this. The `boundary conditions' infinitely far away from the potential barrier(not strictly in the sense of a boundary value problem of a Schr\"odinger differential equation) are the incoming and reflected plane wave states on one side of the barrier and transmitted states on the other side. The transmission and reflection coefficients are then computed by matching the amplitudes at the barrier. In contrast, the case of the IHO is unique in the sense that it is an unbounded potential and the potential has its effect even farther away from the peak. The `eigenvalue' problem thus also provides the scattering situation.  This is seen by studying the behaviour of the states in the position basis far away from the peak of the IHO barrier. 

The asymptotic form of the scattering energy eigenfunctions are give by  \cite{Barton86,Chruscinski03,Chruscinski04}:
\begin{equation}
    \begin{split}
        \chi_E(x \rightarrow -\infty)  &\sim  i \sqrt{\frac{2}{|x|}} {(1+e^{-2\pi E})^{1/2}}e^{-i(\frac{x^2}{4}+E \log|x|+ \phi/2+\pi/4)}\\& -  i \sqrt{\frac{2}{|x|}} e^{-\pi E} e^{ i(\frac{x^2}{4}+E \log |x|+ \phi/2 + \pi/4)}\\
        \chi_E(x \rightarrow + \infty)& \sim \sqrt{\frac{2}{|x|}} e^{ i(\frac{x^2}{4}+E \log |x|+ \phi/2 + \pi/4)} 
    \end{split}
\end{equation}

Here 
\begin{equation}
e^{i\phi(E)}= e^{\pi E/2} \frac{(1+e^{-2 \pi E})^{1/2}}{\sqrt{2 \pi}} \Gamma(\frac{1}{2}-i E)
\end{equation}
From the above one can see that the `plane-wave-like' states in this situation behaves as $e^{\pm ix^2}$. By taking the ratio of the  coefficients of reflected/ transmitted parts to the incident part the reflection and the transmission coefficients can be obtained. Wave packets are constructed by combining these asymptotic forms of the solutions. 

 \subsection{Analytic S matrix: Gamma function}

 Other than the scattering states, the IHO also has associated with it resonant/quasinormal modes. The presence of resonant modes is seen by studying the complex pole structure associated with the S matrix. As clearly seen in Eq. \ref{SMat}, the S matrix is a function of the normalized energy E, which is a continuous variable with support throughout the real number line. To find the complex poles, however, we first analytically extend the scattering matrix to obtain the analytic S matrix. It is poles of this matrix that reveal the presence of resonant modes of the scattering potential.
 
 The analytic S-matrix has played a fundamental role in the history of quantum mechanics and quantum field theory\cite{Eden66,Perelomov} in its role in capturing the essential aspects of a given scattering problem. One cannot extract all the crucial properties of a system, especially in scattering theory, from the real energy eigenstates alone. The poles and zeros in the complex energy plane also manifest as distinct physical phenomena in scattering. One such  aspect is the time-decay in the wave-packet scattering in quantum mechanics. In general the analytic properties of the S-matrix underlies these key features.
 From the above derivations, one can see that the IHO S-matrix  and the energy eigenstates $\langle x\ket{E,+}$ are characterized by Gamma functions $\Gamma(\frac{1}{2}-iE)$. The analytic properties of the Gamma function in the complex energy plane play a key role in determining the IHO phenomena, particularly in determining the existence of temporal decay of wave packets having quantised decay rates.
 
 The Gamma function $\Gamma(z)$ has simple poles  at $z=-n$, where $n=0,1,2...$. Therefore the poles of the IHO scattering problem lie at the imaginary values  :
 \beq
 \tilde{E}_n =-i \bigg( n + \frac{1}{2} \bigg)
 \eeq
 
 These are the resonant poles can also be interpreted as complex energy eigenvalues of the IHO Hamiltonian\cite{Chruscinski03,Chruscinski04}.  In the context of black hole physics, these states which decay in time are known as quasinormal modes and are related ringdown phenomena. In Section \ref{QNM}, we show that these same modes arise as decaying states when wavepackets are scattered across an IHO in a quantum Hall system.


 \section{IHO resonances, Quasinormal modes and Wave packet scattering}\label{QNM}

So far we have discussed about various interesting features in the stationary scattering problem of the IHO. Now we turn our attention towards a detailed analysis of the resonant quasinormal modes of the IHO. In what follows, we present the ladder operator based method to reveal the existence of quantised decay modes, showing that these correspond to purely outgoing/incoming states from infinity. Then we demonstrate that these could be tapped through wave packet scattering off the IHO and also from P\"{o}schl-Teller potential, a realistic counterparts of IHO. Finally we comment on the physical observable that could be accessed through experiments.

\subsection{Resonant/ quasinormal time-decaying states : quantized decay rates}
 
 In the simple harmonic oscillator, existence of a ground state and quantisation of the energy levels is a fundamental manifestation of quantum mechanics. Very much in the same essence, in the IHO existence of time-decay states and the quanitsation of their decay rates are non-trivial manifestation of quantum mechanics. These `quasinormal modes' occur in various contexts from particle decay to modes of perturbations of black holes. 
To understand these modes let us turn to the wavefunctions and the temporal behaviour of the resonant modes of IHO.  To suggestively compare and contrast with operator methods employed in the simple harmonic oscillator, we introduce ladder operators in the `lightcone basis' (setting $\lambda=1$)\cite{Shimbori} 
 \beq
b^{\pm}= (x\pm p)/\sqrt{2}=\frac{1}{\sqrt{2}} \bigg(x \mp i\frac{d}{dx}  \bigg)
\eeq

These operators obey the commutation relations $[b^+, b^-]=-i$,$[b^{\pm}, b^{\pm}]=0$. 
The Hamiltonian takes the form $H =-\{ b^+,b^- \}/2 =(b^+ b^- + b^- b^+)/2$.
 This leads to a relation between ladder operators and the Hamiltonian, similar to that in the Harmonic oscillator:
\beq
[H,b^{\pm}]= \mp i b^{\mp}
\eeq
Equipped with these relations,  one can construct the resonant states of the IHO.
Lets assume that there are a set of states satisfying the condition:
\beq
b^{\mp} w_0^{\pm} = \bigg( \frac{d}{dx}\mp i x \bigg)w_0^{\pm}=0
\eeq
The solutions of this equation are given by
\beq
w_0^{\pm} =B_0^{\pm} e^{\pm ix^2/2}
\eeq

These solutions belong to the `Rigged Hilbert space'\cite{Shimbori,Chruscinski03,Chruscinski04}, which contains additional structure  compared to the regular Hilbert space that allows for states that are not $L^2$ normalizable, but instead are defined in the sense of distributions (like position and momentum eigenstates). Now, one can verify that

\beq
Hw_0^{\pm} = \mp \frac{i}{2} w_0^{\pm}.
\label{0QNM}
\eeq

Therefore, these states can be interpreted as the complex energy eigenstates with eigenvalues $\mp i/2$. One can construct a series of states starting from this state and employing the ladder operators: $(b^{\pm})^n w_0^{\pm}$. The nth states obey the relation:
\beq
H w_n^{\pm}=\tilde{E}_n^{\pm}w^{\pm}_n \quad 
 \tilde{E}_n =-i \bigg( n + \frac{1}{2} \bigg)
\eeq
Reintroducing the strength of the IHO potential $\lambda$
\begin{equation}
    \tilde{\epsilon}_n =-i\hbar \lambda\bigg( n + \frac{1}{2} \bigg)
\end{equation}

The above clearly shows the existence of a time-decaying state with largest decay rate Eq.\ref{0QNM} and a ladder of `excitations' with quantised decay rates, very much in parallel to ground state and quantised energy levels of the SHO. We can see that the scale of the `decay-rate quanta' is set by the strength of the IHO potential $\lambda$. Phenomenologically, the decay rate is an important physical quantity that manifests as life-times in particle decays and quasinormal modes of black hole that carry information about black hole parameters.

 \subsection{Outgoing/Incoming states: Time-decay and probability current flux}
 
One can obtain the wave-functions of the hierarchy of decaying states using the ladder operators in order to study the physical aspects of these states
\begin{equation}
\begin{split}
\bigg[  \frac{1}{\sqrt{2}} &\bigg( \mp i \frac{d}{dx}+x \bigg) \bigg]^n w_0^{\pm}(x) \\
&= B_0^{\pm} \bigg( \frac{\pm i}{\sqrt{2}} \bigg)^n e^{\pm ix^/2} e^{\mp ix^2} \frac{d^n}{dx^n} e^{\pm ix^2},
\end{split}
\end{equation}
we obtain
\beq
w_n^{\pm}(x)=B_n^{\pm} e^{\pm ix^2/2} H_n^{\pm}(x)
\eeq
where $H_n^{\pm}(x) =(\mp)^n e^{\mp ix^2} \frac{d^n}{dx^n} e^{\pm ix^2}$. We refer the reader to Ref. \onlinecite{Shimbori00} for more details about the functions $H_n$ and the normalisation factors $B_n$, which require careful analysis. But as can be seen above, the analysis proceeds in a spirit identical to that determining the ground states and other excited states of the simple harmonic oscillator. 
The stark contrast between the regular stationary states and the resonant states stems from real energy eigenvalues associated with the former. 
The time-dependent wavefunctions for these resonant states are given by
\beq
\psi^{\pm}_n(t,x)=A_n^{\pm} B_n ^{\pm} e^{\mp (n+1/2)t} e^{\pm ix^2} H_n^{\pm}(x).
\eeq 
(Here $A_n$ is the normalisation factor arising form the time-dependant factor \cite{Shimbori00}.)
The immediate observation is that these states decay or grow in time. 
The associated probability densities are given by\cite{Shimbori00}
\beq
\rho^{\pm}_n(t,x)=|A_n^{\pm}|^2 |B_n ^{\pm}|^2 e^{\mp (2n+1)t} H_n^{\mp}(x)H_n^{\pm}(x)
\eeq
 and the currents are given by
 \beq
 \begin{split}
 j_n^{\pm}(t,x) = \pm |A_n^{\pm}|^2 |B_n ^{\pm}|^2 e^{\mp (2n+1)t} (xH_n^{\mp}(x)H_n^{\pm}(x) \\ \pm 2n \text{Im}[H_n^{\mp}(x)H_{(n-1)}^{\pm}(x)])
 \end{split}
 \label{Eq:RCurrents}
 \eeq
 As with ordinary states, these states satisfy the continuity equation:
 \beq
 \frac{\partial}{\partial t}\rho^{\pm}_n (t,x) + \frac{\partial}{\partial x} j^{\pm}_n(t,x) =0
  \eeq

Finally the asymptotic behaviour of the currents in Eq.\ref{Eq:RCurrents} is given by
\beq
j_n^{\pm}(t,x) \approx  \pm e^{\mp (2n+1)t} x^{2n+1}.
\eeq

We see from this form that the probability density decays in time but current conservation ensures that this decay manifests as a finite current that goes out to infinity(thus the finite value of the wavefunction at infinity)

Therefore, the resonant modes correspond to  purely incoming states or purely outgoing states in one direction(left or right). They thus require having finite amplitude at infinity. Such states do not belong to the regular Hilbert space. One needs to enlarge the Hilbert space to so called `Rigged Hilbert space' \cite{Shimbori, Chruscinski03}. Even in the simplest problem of scattering against a barrier, such as a square potential, one uses plane waves that are not normalizable and thus do not strictly belong to the regular Hilbert space.  As we have already remarked, these states are part of the stationary scattering states of the problem. But in the discussion of resonant states, the rigged Hilbert space furnishes states necessary for a dynamical scattering scenario where the amplitude decays in time and has finite amplitude far away from the barrier.  
 The purely outgoing behavior is intricately related to the time decay of the wavefunctions and probabilities. Even in a model as simple as the IHO, one can see that in a scattering problem, enlarging the set of allowed boundary conditions can lead to time decay behaviour. Such behavior is captured in the resonant pole structure of the problem and can be attributed to the complex energy eigenvalues.
From the above expression, one can also see that the `decay rates' of the wavefunctions are quantized as $(n+1/2)$, much like the  bound state energies of the simple harmonic oscillator. One can trace the `zero-point' factor of $1/2$ in the SHO derivation(as well as in IHO) from quantum fluctuations associated with the commutation relation $[X,P]=i$. This quantization and existence of a bound on the decay rate is a fundamental manifestation of quantum mechanical scattering in the IHO.
\subsection{IHO resonances and wave-packet scattering}
\label{WaveScat}

We have shown that the IHO is the effective Hamiltonian for a saddle potential in the LLL. Hence, the resonant spectrum of the IHO must also manifest in the quantum Hall system and any other system that hosts the IHO. Here, we pinpoint the effects of such resonances and how they become manifest in wave-packet scattering. 

To first briefly recapitulate the details of the previous sections, it is important to draw attention to the analytic structure of the scattering matrix of the IHO. On analytically extending the matrix to the complex energy plane, we see that the Gamma function gives the S matrix an infinite number of poles in the lower half plane. That is, the Gamma function $\Gamma(x)$ has poles at $x=-n, n\in \mathbb{I^+}$. Therefore, the S-matrix has poles at $E_n=-i(n+\frac{1}{2})$. These poles are the source of resonant states in the scattered wavepacket. 

Equivalently, one might define resonant states as those which carry a complex energy corresponding to the poles of the S matrix. That is, a resonant state $\psi_n$ satisfies
\beq H_{IHO}\psi_n=\tilde{E}_n \psi_n\eeq
leading to a discrete complex spectrum to the IHO. The presence of complex energy eigenstates is no cause for alarm, and does not violate the hermiticity of the Hamiltonian. Instead, we have states that decay in time, but grow with distance, such that they have a finite amplitude at the boundary of the system. Thus, resonant states describe scattering experiments with a well defined outgoing current at the boundary of the system. For this reason, they are useful in describing particle decay processes \cite{Bohm}. They may be observed by scattering a wavepacket against the IHO. We demonstrate this analytically by picking a Gaussian wavepacket.


 To capture the effect of resonances which are specific poles of the scattering matrix, we need to consider an incident wavepacket composed of scattering states of different eigen-energies following the notation in Ref. \onlinecite{Barton86}, we have
 \begin{equation}
 \Psi_i=  \frac{i}{\sqrt{|x|}} \int dE \tilde{f}(E) e^{-i(\frac{x^2}{4}+E \log|x|+ \phi/2+\pi/4)} e^{-iEt}.
 \end{equation}
 The envelope function $\tilde{f}(E)$ is peaked near $E_0$ and is normalised as $2 \pi \int dE |\tilde{f}(E)|^2=1 $.
The incident wave can thus be rewritten as
\begin{equation}\begin{split}
\Psi_i = i \sqrt{\frac{1}{|x|}} &e^{-iE_0t-i \Phi_0}\times \\
&\int dE [\tilde{f}(E)e^{-i(\phi-\phi_0)/2}]e^{-i(E-E_0)(t+log|x|)}.
\end{split}\end{equation}

Here $\Phi_0= E_0\log|x|+x^2+ \phi_0/2 +\pi/4$ and $f_i(E)= \tilde{f}(E)e^{-i(\phi-\phi_0)/2}$
We choose the wave-packet to be a Gaussian in energy centered around $E_0$:
\begin{equation}\begin{split}
\int_{-\infty}^{+\infty} dE f_i(E)&e^{-i(E-E_0)t}\\ &=\int_{-\infty}^{+\infty} dE (\frac{1}{2 \pi^{3/2})^{1/2}\Delta}e^{\frac{|E-E_0|^2}{2 \Delta^2}} e^{-i(E-E_0)t}\\&= (\frac{\Delta}{\pi^{1/2}})^{1/2} e^{-t^2 \Delta^2/2}
\end{split}\end{equation}
Therefore, the incident wavepacket may be written as
 \begin{equation}
 \Psi_i= i e^{-i E_0t-i\Phi_0}\sqrt{\frac{1}{|x|}}\frac{\Delta^{1/2}}{\pi^{1/4}}e^{-\Delta^2 (t+ \log|x|)^2/2}.
 \end{equation}


 The reflected wave-packet can be written down from terms in the scattering matrix. This is then given by
 \begin{equation}\begin{split}
   \Psi_R &= -ie^{-iE_0t+i\Phi_0} \sqrt{\frac{1}{|x|}}\int dE f_r(E)e^{-i(E-E_0)(t-\log|x|)}\\
  f_r(E)&=\frac{1}{\sqrt{2\pi^{3/2} \Delta}} e^{\frac{-(E-E_0)^2}{2\Delta^2}}  \frac{e^{-\pi E}e^{i(\phi- \phi_0)}}{ (1+e^{-2\pi E})^{1/2}}   
 \end{split}\label{psiri}\end{equation}
 The integrand $\mathcal{F}_r$ can be rewritten by making the substitution
 \begin{equation}
 e^{i\phi}=e^{\pi E/2} \frac{(1+e^{-2 \pi E})^{1/2}}{(2\pi)^{1/2}} \Gamma (\frac{1}{2}-iE),
 \end{equation}
 to obtain a form that makes its pole structure more apparent. Thus, we have
 \begin{equation}\begin{split}
\mathcal{F}_r&= \int dE f_r e^{-i(E-E_0)(t-\log |x|)}\\&= \int dE \frac{e^{-\frac{(E-E_0)^2}{2\Delta^2}-\frac{ \pi(E-E_0)}{2}}}{\sqrt{2 \pi^{3/2}\Delta}} \frac{\Gamma(\frac{1}{2}-iE)}{\Gamma(\frac{1}{2}-iE_0)}e^{-i(E-E_0)(t-\log|x|)}
 \end{split}\end{equation}

 Now, using standard methods of scattering theory, we can extend the above integral into complex plane. We see from above that the Gamma function within  the integral has a pole in the lower half energy plane. To access the lower half plane, we consider the times $t> \log|x|$.  The poles of the Gamma function $\Gamma(x)$ are at $x=-n; n=0,1,2$. Therefore, the poles of $\Gamma(1/2-iE)$ are given by
 \begin{equation}
 E_n=-i(n+\frac{1}{2}).
 \end{equation}
 The corresponding residue of the integral for  the reflection amplitude is then given by
 \begin{align}
 \text{Res}&[\mathcal{F}_r; E= -i(n+1/2)]=\frac{e^{-(-i(n+\frac{1}{2})-E_0)^2}/(2\Delta^2)}{\sqrt{2 \pi^{3/2}\Delta}} \nonumber\\&e^{- \pi(-i(n+\frac{1}{2})-E_0)/2}\frac{(-1)^ne^{-i(-i(n+\frac{1}{2})-E_0)(t-log|x|)}}{\Gamma(\frac{1}{2}-iE_0) n!}. 
 \end{align}
Extracting the dominant temporal and spatial aspects of the reflected form from the expression above , we have the form(considering the contribution of one pole and reintroducing the scale of the IHO potential $\lambda$)
 \begin{equation}
\Psi_r \sim e^{-t/2 \lambda} e^{log\sqrt{|x|}}\label{psirf}
\end{equation}
This indicates that the reflected wave decays exponentially in time and has finite amplitude at large x. The decay rate is determined by $\lambda$. While we have shown here that the manifestation of resonances arising from the pole structure of the scattering matrix, resonances can also be studied as states with complex eigen-energies \citep{Chruscinski03, Chruscinski04}. As an important application,traits of these decaying solutions are characteristic of the quasinormal modes occurring in the context of black holes\cite{Vishveshwara}. We shall give a detailed description of black hole QNMs in a later section.

 

\subsection{P\"{o}schl-Teller potential}

One issue with the IHO in realistic situations, such as the quantum Hall point contact is that is an unbounded potential and hence, is not physical.
 Instead, one may choose a bounded variant of a peaked scattering potential, and still see the occurrence of resonant poles. To this effect, we choose the hyperbolic family of P\"{o}schl-Teller (PT) potentials. These potentials are exactly solvable in 1D, and thus, their scattering properties may be exactly derived. Here, we demonstrate the appearance of resonant states in bounded scattering potentials \cite{cevik} for the class of PT potentials whose Hamiltonian is give by

\begin{align}
    H=-\frac{\hbar^2}{2m}\frac{d^2\psi}{dx^2}-\frac{\hbar^2}{2m}\frac{\alpha^2\lambda(\lambda-1)}{\cosh^2\alpha x}\psi\label{ptpot}
\end{align}

\begin{figure}
    \includegraphics[width=0.5\textwidth]{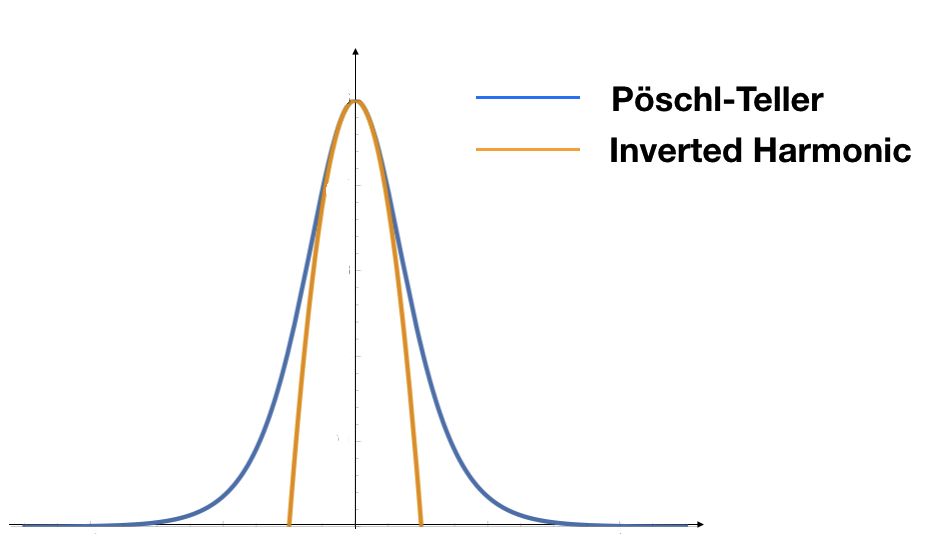}
    \caption{P\"{o}schl-Teller is the bounded variant of a scattering potential peak, which tapers off at large distances. The IHO is an approximation to P\"{o}schl-Teller potential close to the peak. }
    \label{poschl}
\end{figure}
Without loss of generality, we can set $\alpha=1$.  To obtain barrier potentials, the parameter $\lambda$ has to take values $\lambda=\frac{1}{2}+il, l>0$. The eigenstates of the Hamiltonian take the form\cite{cevik}

\beq \begin{split}
 &\psi(x)=  A R(x)^{ik/2}    \textnormal{}_2F_1\big(\lambda,1-\lambda;ik+1;\frac{1+\tanh  x}{2}\big)\\ &+B2^{ik}R(x)^{-ik/2}\textnormal{} _2F_1\big(\lambda-ik,1-\lambda-ik;1-ik;\frac{1+\tanh  x}{2}\big)
 \end{split}
\eeq
where $R(x)=\frac{(1+\tanh x)}{(1-\tanh  x)}$, and A and B are constants. The functions of the form $_2F_1(a,b;c;x)$ are hypergeometric functions as defined in Ref.~\onlinecite{Abramowitz}. This scattering potential also has a resonant pole structure that can be gleaned from the scattering matrix. The components of the scattering matrix can be found as usual by writing down the asymptotic forms of the eigenfunctions. The asymptotic forms of the hypergeometric function have been extensively studied \cite{NIST:DLMF}, and as expected for a bounded potential, the asymptotic form of the eigenfunction is proportional to plane waves. Thus, one can obtain expressions for the transmission and reflection coefficients for $\lambda=\frac{1}{2}+il$,
\beq T=|t|^2=\frac{\sinh^2 \pi k}{\cosh^2 \pi l+\sinh^2 \pi k }  \eeq
\beq R=|r|^2=\frac{\cosh^2 \pi l}{\cosh^2 \pi l+\sinh^2 \pi k }  \eeq

The resonances, defined as poles of $T$ and/or $R$ in the complex $k$ plane, belong to two sets of points in the complex $k$ plane, given by
\begin{align}
    k_1(n)&=l-i(n+\frac{1}{2})\\
    k_2(n)&=-l-i(n+\frac{1}{2})
\end{align}
Here, the $k_1(n)$ series of poles correspond to decaying modes while $k_2$ corresponds to growing modes. 

Asymptotically, the PT potential goes to zero for large $|x|$, and thus, far away from the origin, we may approximate any wavepacket as a superposition of plane waves. Thus, we can take $\phi_t(x)\sim \int dk S(k)e^{\pm ikx}e^{iEt}$.  We can write a density function $\rho \sim \phi_t(x)^\dagger\phi_t(x)$ for the transmitted wavepacket as 
\beq\rho(x,t)=\int dk T(k) e^{-2x \textnormal{Im}(k)}e^{2t \textnormal{Im}(E)}. \eeq 
We can perform this integral by picking an appropriate contour on the complex-k plane. While more complicated than the choice involving the lower half plane of the complex E plane in the case of the IHO, the contour integral can still be evaluated analytically\cite{Perelomov}. One thus obtains the form
\beq\rho(x,t)=\sum_{n=0} \frac{\coth \pi l}{2} e^{(2n+1)x}e^{-\hbar^2 tl(2n+1)/2m} \eeq
In general, it can be shown that a bounded barrier potential, or a barrier potential with a finite region of support, has QNMs that take the form $\psi\sim e^{\Gamma(x/v-t)}$, where $\Gamma$ is the decay rate, and $v$ is an effective velocity that may be determined from the semi-classical equations of motion of the wavepacket \cite{Bohm}. Therefore, the reflected and transmitted wavepackets will have a decaying component. We see that the PT potential follows this pattern as well. 

As with the IHO, we can define ladder operators that act as raising and lowering operators for the resonant modes of the PT potentials. That is, we define $K^{-}_{j,n}$ and $K^{+}_{j,n+1}$ such that 
\begin{equation}
    K^{-}_{j,n}:\varphi_{j,n}\rightarrow\varphi_{j,n-1}\textnormal{, } K^{+}_{j,n+1}:\varphi_{j,n}\rightarrow\varphi_{j,n+1}
\end{equation}

where $\varphi_{j,n}$ corresponds to a resonant mode (or residue corresponding to a resonant pole) at $k_j(n), j\in 1,2$ for each series of poles. These operators are given by 
\begin{equation}
    \begin{split}
        K^{-}_{j,n}=-\cosh x \partial_x +ik_j(n)\sinh x\\
        K^{+}_{j,n+1}=\cosh x \partial_x +ik_j(n)\sinh x
    \end{split}
\end{equation}
These operators also span the $\mathfrak{sp}(2,\mathbb{R})$ algebra along with the operator $K^0_{j,n}$ which acts diagonally on $\varphi_{j,n}$\cite{cevik}.

\subsection{Physical Observables}

Time evolution of a quantum mechanical system that shows decaying behaviour has been studied in many contexts, such as nuclear radioactivity, dynamical systems, quantum chaos and tunnelling of coherent states. Often, it is necessary to make measurements of the time decaying state that clarify the nature of the decay itself - say, to differentiate between exponential and power law decays. To this end, we adopt the Fock-Krylov method \cite{Ram_rez_Jim_nez_2019} of specifying survival probabilities and elucidate the nature of resonant decay in our system. 

Consider a quantum mechanical system with an applied potential $V(x)$ that goes to zero for $|x|>R$. We are interested in the nature of states $\psi(x)$ that are solutions to the Schrodinger equation for $|x|<R$, and then decay out to the space beyond $R$. The survival amplitude, and survival probability, are defined as
\begin{align} A(t)=\braket{\psi(0)|\psi(t)}&=\int\limits_{-R}^R dx \psi^*(x,0)\psi(x,t)\\ S(t)&=|A(t)|^2 \end{align}
The survival probability at $t>0$ is a measure of the probability of finding the particle in its initial state at time. A very similar quantity called the non-escape probability can also be defined as 
\beq P(t)=\braket{\psi(t)|\psi(t)}=\int\limits_{-R}^R dx \psi^*(x,t)\psi(x,t). \eeq
This quantity is a measure of the probability that the state, at time t, remains within the region $|x|<R$. In the case of exponential decay, the time dependent component of these probabilities would go as a combination of $\sim e^{-\Gamma_i t}$. A logarithmic plot of these probabilities is a measure of the dominant decay rate $\Gamma_i$, and hence, the poles of the scattering matrix corresponding to the barrier potential.

The Fock-Krylov approach adapts these definitions to wavepackets (therefore not eigenstates) that may be scattering or tunnelling through the applied potential, since in a continuous spectrum, the post-tunnelling state is not an eigenstate of the Hamiltonian. Consider an initial wavepacket of the form
\beq \ket{\psi(0)}=\int dE\textnormal{ } a(E)\ket{E} \eeq. Therefore, we have 
\begin{equation}\begin{split} A(t)&=\int dE\textnormal{ } |a(E)|^2 e^{-iEt}.
\end{split}\end{equation}
Due to the presence of resonant poles with nonzero imaginary part, we deduce that $A(t)\sim e^{-\Gamma_n t}$ after appropriate contour integration, where $\Gamma_n$ are the imaginary parts of the resonant poles at $E=E_n$.

\begin{figure}
    \centering
    \includegraphics[width=0.4\textwidth]{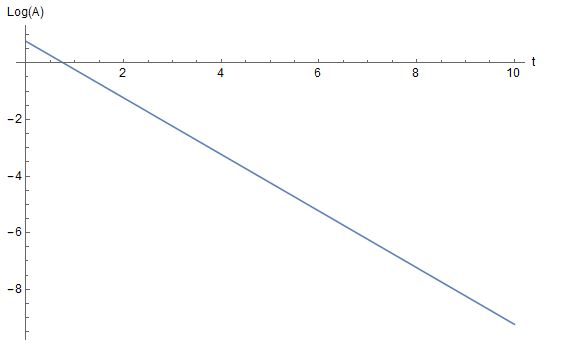}
    \caption{Logarithmic plot of the survival amplitude as defined in Eq.\ref{Ampl}. For a single pole, the plot is linear as seen in this figure, with the slope giving the decay rate of the reflected wavefunction. In the presence of higher order poles, the plot would no longer be linear, and the slope indicates instantaneous decay rate.}
    \label{SurvAmp}
\end{figure}

For resonant decay in the case of the IHO, there is no boundary at $|x|=R$ from which the decay process is observed. But realistically, if the IHO is applied to the system by means of, say, a quantum point contact, then this QPC itself does not have infinite support. So one may reasonably assume some large $R$ beyond which the IHO is not applied. Alternatively, from the form of the resonant modes, we see that the decay process at any given point $x$ is only observable after a time $t(x)=\log|x|$. This time can serve as the starting point for observation of the decay at any given distance. That is, we modify the definition of survival amplitude to 
\beq A(t)=\int\limits_{-\infty}^\infty dx \psi^*(x,t'(x))\psi(x,t'(x)+t)\label{Ampl}.\eeq For IHO resonances, the integrand goes as $\sim x^{-1}$ and a principle value can be obtained when the integral is curtailed to a large R. For a single dominant pole, as shown in Eq. \ref{psirf}, the survival amplitude as well as the survival probability are just a decaying exponential function in time, and their logarithmic plot would be a straight line as seen in Fig.\ref{SurvAmp}. The presence of higher order poles would lead to curves of with a different value of effective decay rate at every instant of time. 
The decay rate may be experimentally determined through time-of-flight measurements in electron optics setups, particularly like those described in Ref. \cite{Kataoka}. Electrons are released via a single-electron pump as wavepackets centered at a given energy. These wavepackets travel along the edge of a depleted quantum Hall setup and approach a QPC.  Detectors on the reflected (or equivalently, transmitted) region at a specified distance can detect the scattered wavepacket, we can calculate overlap quantities of the general form
\beq \braket{\psi_i(0)|\psi_{r}(t)}=\int dE |a(E)|^2r(E)e^{-iEt} \eeq
which can show the presence of resonances. 
 
 We have reviewed the physics of inverted Harmonic oscillator and shown its realisation in a quantum Hall system under the influence of an external potential. Now we shall turn to review the basics of Hawking-Unruh effect. We highlight the Rindler Hamiltonian as the fundamental object underlying quantum mechanics near an event horizon of a black hole and show that the physics of IHO directly parallels the key aspects of it.

 \section{Rindler Hamiltonian, Hawking-Unruh effect and IHO physics}\label{RindHam}

 Here, we build up to connecting the IHO concepts discussed in previous sections. We first conceptually introduce black holes, horizons, and light cones. We then more formally elaborate on these concepts for the unfamiliar reader as well as to delineate our approach in drawing parallels in the quantum Hall setting. Specifically, we present the notion of a Rindler observer as one in an accelerating frame and show that time translations in the Rindler frame correspond to Lorentz boosts in flat spacetime (i.e boost is the generator of time translation/ a Hamiltonian). This setup has a direct manifestation in the time evolution of the quantum mechanical states in the Rindler spacetime. The dynamics generated by the Rindler Hamiltonian, in this outlook, is at the essence of the Hawking-Unruh effect. We show that the dynamics of IHO parallels that of the Rindler Hamiltonian. The scattering amplitudes of the IHO exactly match the Bogoliubov coefficients that appear in the time evolution operator for the Rindler Hamiltonian. This parallel directly leads to the effective thermal form of the tunneling probability in a saddle potential. When applied to the quantum Hall problem, we will see that this lets us explain the thermal form of the tunnel conductance through a point contact, as well as the relationship between Hall viscosity and Wigner rotations. We remark here that the thermal parallel is purely formal; tunneling still corresponds to a zero temperature quantum process in which the strength of the scattering potential appears a temperature-like factor, as shown in Eq.\ref{Tval}.  In elaborating on these concepts  more technically below, we make use of the term `Hawking-Unruh effect' to refer to the thermal nature of quantum mechanical states with respect to Rindler time-evolution.

 \subsection{Rindler wedge, Rindler Hamiltonian and Lorentz boost}

To define Rindler space and the Rindler Hamiltonian, first consider 1+1-dimensional Minkowski (flat) space-time $(t,x)$ described by the metric $ds^2=dt^2-dx^2$. There is a lightcone that determines the causal structure of the spacetime. Events at $x=0$ can have causal connection with time-like (also known as light-like) separated events i.e events with $ds^2\geq 0$; such events lie within or on the lightcone. Other regions in $x>0$ and $x<0$, the wedges`under' the lightcone, are said to be `space-like'. Events in these wedges cannot have signals propagating beyond the lightcone and hence events in the space-like wedges cannot affect events in the lightlike wedges. Note that Lorentz boosts cannot move an observer outside of a space-like or light-like wedge.

\begin{figure}
   \centering
\includegraphics[width=0.4\textwidth]{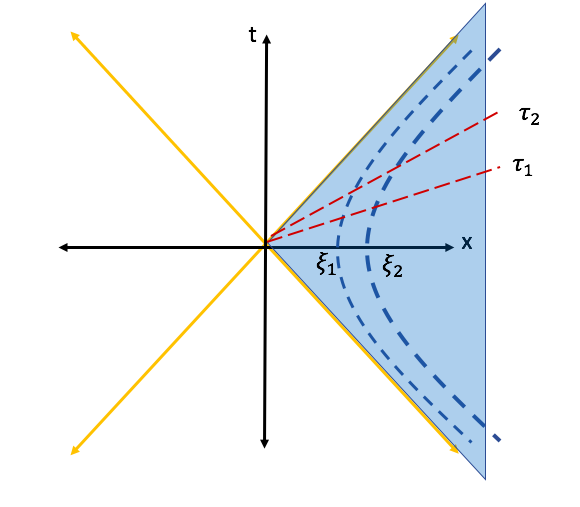}
\caption{Spacetime diagram for Minkowski spacetime and the right Rindler wedge. The lightcone structure in the Minkowski spacetime bifurcates the spacetime into spacelike and time like regions. A family of observers with constant acceleration are indicated by hyperbolic trajectories. These observers are confined to the spacelike region shaded in blue. This region is known as the `right Rindler wedge', which can be described in terms of co-ordinates ($\tau,\xi$).  The constant time slices are shown by slanted lines. The translation corresponds to hyperbolic rotation in the Minkowski space.}
\label{fig:Rindler}
\end{figure}

The spacetime of interest, the `Rindler spacetime' is given by the metric:
\beq
ds^2 =e^{2\kappa \xi} ((d\tau)^2-(d\xi^1)^2).
\label{RindMet}
\eeq
The co-ordinates have the range: $- \infty <\tau< \infty$,  $-\infty < \xi< + \infty$.
This metric can be obtained from the Minkwoski metric through the follwing co-ordinate transformation-
\beq
t=\frac{e^{\kappa \xi}}{\kappa}\sinh{\kappa \tau} \quad x=\frac{e^{\kappa \xi}}{\kappa} \cosh{\kappa \tau}
\label{Eq:MinkToRindler}
\eeq
As shown in Fig.\ref{fig:Rindler}, the co-ordinates $(\tau,\xi)$ cover only the wedge $x>|t|$ in Minkowski spacetime. This is called the `right Rindler wedge'. The `left Rindler wedge', which covers the region $x<|t|$ in the Minkowski spacetime can be obtained by changing the signs on the right hand sides of the above transformation. This is equivalent to doing a time reversal transformation, followed by a spatial reflection.

 The lines of constant $\tau$ value correspond to constant proper time slices and are lines of constant slope in the $(t,x)$ plane, as shown in Fig.\ref{fig:Rindler}.
 The curves of $\xi = \text{constant}$ are hyperbolae $x^2-t^2=e^{2\kappa \xi}/\kappa$ as shown in Fig.\ref{fig:Rindler}. These correspond to trajectories of uniformly accelerating observers. We shall call these observers as `Rindler observers'. Another useful form of expressing the Rindler metric is in terms of the co-ordinates $(\rho, \tau): \rho =e^{\xi}, \tau = \tau$(setting $\kappa=1$):
\beq
ds^2=\rho^2 d\tau^2 -d\rho^2
\eeq
This representation gives an angular interpretation for the time-like co-ordinate $\tau$.

 As can be seen, the hyperbolic trajectories of the Rindler observers asymptote to the light cones at infinities and in fact do not cross them. Therefore, the lightcones act as horizons for Rindler observers in both the wedges. In fact, the right and left Rindler wedges are causally disconnected as events in one cannot causally affect the events in the other wedge. in this sense the lightcone naturally partitions the Minkwoski spacetime into two causally separated parts. Each wedge can be considered as a spacetime in its own right and is often called a `Rindler universe'. 

Now let us consider a crucial fact about time evolution and dynamics within a Rindler wedge: 
{The time translation $\tau\rightarrow \tau+\beta$ in the right Rindler wedge is a Lorentz boost having rapidity $\beta$ with respect to the Minkowski spacetime. }
   
   One of the simplest ways to see this in the Rindler spacetime is to go to so-called `lightcone co-ordinates' $u= t-x, v=t+x$. The Lorentz boost along x-direction can be expressed as a hyperbolic rotation: $t \rightarrow t \cosh \beta + x \sinh \beta$, $x \rightarrow t \sinh \beta +x \cosh \beta $, where $\beta$ is called the {\it rapidity} parameter and is related to the velocity `v' of the boost through $\tanh{\beta}=v/c$ . If we switch to lightcone co-ordinates , the boost looks much simpler : $u \rightarrow u e^{\beta}$, $v \rightarrow v e^{-\beta}$. 
   
   Now relating the  lightcone co-ordinates to the Rindler co-ordinates, we obtain
   \beq
   u = e^{\xi +\tau};  \quad v=e^{\xi-\tau}.\label{LC} 
   \eeq
   One can immediately see from the above that a time translation $\tau \rightarrow \tau+\beta$ is a boost in the lightcone co-ordinates.  
   
   We have particularly highlighted this fact about the time-translation in the Rindler wedge as it determines the Hamiltonian that acts on the quantum mechanical modes and is important in the understanding of the Hawking-Unruh effect. 
   

\subsection{The Hawking-Unruh effect and structural parallel to IHO} \label{BoostIHO}

We are now equipped to draw the parallel between quantum mechanics near an event horizon that leads to Hawking-Unruh effect and the physics of the IHO. The parallel we demonstrate in this work is that of the identical structure shared by the quantum mechanical modes in the Rindler wedges and the scattering states of the IHO.  We also comment on the underlying isomorphisms in the algebras of the symmetries in the two platforms.

The simplest derivation of the Hawking-Unruh effect for a non-interacting scalar field proceeds as follows \cite{Mukhanov,Davies}. Hawking radiation \cite{Hawking} is the phenomenon of emergence of thermal radiation from the black hole event Horizon as observed by a far away observer and is a manifestation of the fluctuations of the quantum fields near the event horizon. The Unruh effect\cite{Unruh76} is the emergence of a thermal bath for a uniformly accelerating observer in a Minkowki spacetime. In this work we use the collective term `Hawking-Unruh' effect for the above two phenomena {The Hawking-Unruh effect is the phenomenon where the vacuum state for a quantum field theory in Minkowski spacetime restricted to the right Rindler wedge is a thermal state with respect to Rindler time evolution}

Consider the scalar field $\phi$ in the Minkowski spacetime. We are interested in comparing the `particle' content of the field in the Minkowksi and Rindler spacetime.  ``Particles" are  defined as positive frequency modes of a given field. But the positive frequency is defined with rest to the proper time of the observer. As we saw, the time co-ordinate in the Minkowski ($t$) and the accelerating frame ($\tau$) are related in a non-trivial way. The Rindler time translation is a Lorentz boost. As a result, comparing the positive frequency modes results in a Bogoliubov  transformation operators associated with the hyperbolic transformation of the boosts. Thus the notion of particle changes when we switch frames and  leads to the Hawking-Unruh effect \footnote{Note that we are restricting the discussion of Hawking-Unruh effect to the approaches taken in \onlinecite{Hawking,Unruh76} in terms of `particle content'. This is the simplest approach but also the most restricted one confined to non-interacting theories.}.

The equation of motion for the scalar field is given by the wave equation
\begin{align}
    \Box \phi =\frac{1}{\sqrt{-g}}\frac{\partial}{\partial x^{\nu}} \bigg( g^{\mu \nu} \sqrt{-g} \partial_{\mu} \phi \bigg)=0
\end{align}
The metric for the Minkowski and Rindler spacetimes are conformally equivalent through Eq. \ref{Eq:MinkToRindler}. So the solutions to the equations of motion are plane waves in each space. That is, the solutions in the Minkowski space are given by 
\begin{equation}\begin{split}
 \phi \propto e^{\pm i\omega u}=e^{\pm i \omega (t-x)} \\
 \phi \propto e^{\pm i\omega v}=e^{\pm i \omega (t+x)} 
\end{split}\end{equation} 
while the  solutions in the Rindler space are given by
\begin{equation}\begin{split}
 \phi \propto e^{\pm i\Omega \tilde{u}}=e^{\pm i \Omega (\tau-\xi)}
 \phi \propto e^{\pm i\Omega \tilde{v}}=e^{\pm i \Omega (\tau+\xi)}\label{RModes}
\end{split}\end{equation}
Rewriting the Rindler modes in Eq. \ref{RModes} using the lightcone coordinates in Eq. \ref{LC}, we see that they take the form $|u|^{\pm i\Omega}$ or $|v|^{\pm i\Omega}$. More specifically, the Rindler modes in the right wedge $x>|t|$ take the form
\beq 
\Theta(-u)(-u)^{i\Omega}, \quad  \Theta(v)(v)^{-i\Omega} \label{RW}
\eeq 
and their complex conjugates. In the left wedge $x<|t|$, they are
\beq 
\Theta(u)(u)^{-i\Omega}, \quad  \Theta(-v)(-v)^{i\Omega} \label{LW}
\eeq
and their complex conjugates. In comparing the modes in Eqs. \ref{RW} and \ref{LW} with  Eqs.(\ref{Eq:Mode1}-\ref{Eq:Mode4}), we see that the Rindler modes are in fact incoming/outgoing eigenstates of the IHO.

There is a crucial difference in the factor of $1/2$ in the exponent of the Rindler/IHO modes: $i\Omega=iE-\frac{1}{2}$. This factor of half provides the correct measure of integration for the normalisation of the IHO eigen modes. The same factor of half also shows up as a difference in a crucial sign in the final thermal distribution we obtain in the two cases. 

 Connecting back to our discussion of Lorentz boost being the Hamiltonian(generator of time translation) in the Rindler wedge, the Rindler modes obtained here are in fact eigenmodes of the boost generator. The boost generator in the Minkowski space-time is of the form\cite{Padmanabhan19,Bliokh}:
 \beq
 -i(x \frac{\partial}{\partial t}+t\frac{\partial}{\partial x}),
 \eeq
 which is a hyperbolic rotation mixing space and time. The eigenmodes are $|-t\pm x|^{-i\Omega}$. This reduces to the Rindler modes in the lightcone co-ordinates.  We obtained the Rindler modes as solutions to the wave equation, which is a relativistic d'Almbertian (i.e. $p^2-E^2 \sim \partial^2_\tau-\partial^2_\xi$). The boost generator commutes with the d'Alembertian (because of the Poincare algebra) and thus the two operators share the same eigen modes\cite{Bliokh}.  Thus by realising the IHO in a physical setup like the quantum Hall system, we are directly accessing the boost eigensystem, which lies at the heart of Hawking-Unruh thermality.

Let us proceed further to  quantize the scalar field by writing its mode expansions in terms of plane waves in both spacetimes in the following way - 
 \begin{equation}
 \hat{\phi} =\int_0^{\infty} \frac{d \omega}{\sqrt{4\pi \omega}}[e^{-i\omega u}\hat{a}_{\omega}+ e^{i \omega u} \hat{a}^{\dagger}_\omega]+\textnormal{(left moving)}
 \end{equation}
 
  \begin{equation}
 \hat{\phi} =\int_0^{\infty} \frac{d \Omega}{\sqrt{4\pi \Omega}}[e^{-i\Omega \tilde{u}}\hat{b}_{\Omega}+ e^{i \Omega \tilde{u}} \hat{b}^{\dagger}_\Omega]+\textnormal{(left moving)}\label{ME2}
 \end{equation}
 where the operator $\hat{a}_{\omega}$ acts on the Minkowski vacuum $\ket{0_M}$ and $\hat{b}_{\Omega}$ acts on the Rindler vacuum $\ket{0_R}$. The modes in the right and the left wedges are themselves not complete enough to do the mode expansion of $\phi$ but taken together, they can be used to quantize the field $\phi$ with creation and annihilation operators $b_{\Omega}, b_{\Omega}^{\dagger}$\cite{Davies}. 

 The two operators are related by Bogoliubov transformations \cite{Mukhanov,Davies}.
 \begin{equation}
 \hat{b}_{\Omega}=\int_0^{\infty} d\omega [\alpha_{\Omega \omega} \hat{a}_{\omega} - \beta_{\Omega \omega} \hat{a}^{\dagger}_{\omega}]\label{rel}
 \end{equation}
 
 The Bogoliubov coefficients, by definition, are required to satisfy
 \begin{align}
     \int_0^\infty d\omega (\alpha_{\Omega,\omega}\alpha^*_{\Omega',\omega}-\beta_{\Omega,\omega}\beta^*_{\Omega',\omega})=\delta(\Omega-\Omega')
 \end{align}
 Plugging Eq. \ref{rel} into Eq. \ref{ME2}, we immediately find that
 \begin{align}
     \frac{1}{\sqrt{\omega}}e^{-i\omega u}=\int_0^\infty \frac{d\Omega'}{\Omega'}(\alpha_{\Omega',\omega}e^{-i\Omega'\tilde{u}}-\beta^*_{\Omega',\omega}e^{i\Omega'\tilde{u}})
 \end{align}
 Multiplying both sides by $e^{\pm i\Omega\tilde{u}}$, and integrating over $\tilde{u}$, we see that the integral is a Mellin transform(See Eq.  \ref{mellin}) , which yields the coefficients $\alpha_{\Omega \omega}$ and $\beta_{\Omega \omega} $
 
  \begin{equation}
 \alpha_{\Omega \omega}, \beta_{\Omega \omega} = \pm \frac{1}{2 \pi \kappa}\sqrt{\frac{\Omega}{\omega}}e^{\pm \frac{\pi \Omega}{2\kappa}}exp \bigg( \frac{i\Omega}{\kappa}ln\frac{\omega}{\kappa}\bigg)\Gamma(-i\Omega/\kappa)\label{BT}
  \end{equation}
  Up to numerical factors, we see that the Bogoliubov coefficients in Eq \ref{BT} are identical to the terms in the scattering matrix in Eq. \ref{SMat}, after making the all important conversion of $i\frac{\Omega}{\kappa}$ to $iE-\frac{1}{2}$:
   \begin{equation}
  \hat{\mathcal{S}}= \frac{1}{\sqrt{2 \pi}} \Gamma\bigg( \frac{1}{2}-iE  \bigg)  \left(\begin{array}{cc}
 e^{-i\pi/4}e^{-\pi E/2}&
  e^{i\pi/4}e^{\pi E/2}\\
   e^{i\pi/4}e^{\pi E/2}& e^{-i\pi/4}e^{\pi E/2}\end{array}\right)
  \end{equation}
  
  That is, the coefficients relating the incoming and outgoing states scattering off an IHO are the same as the Bogoliubov coefficients relating Rindler modes to Minkowski modes.
  The meaning of these coefficients is further seen in explicitly writing down the average density of the number of particles in the Minkowski vacuum as seen by the Rindler observer in the following way - 
  \begin{equation}\begin{split}
  n_\Omega&=\frac{ \langle \hat{N}_{\Omega} \rangle}{V}=\\
  &=\frac{1}{V}\bra{0_M} \hat{b}_{\Omega}^{\dagger} \hat{b}_{\Omega} \ket{0_M} =\frac{1}{V}\int d\omega |\beta_{\omega \Omega}|^2\\
  &=\frac{1}{1-exp(-2\pi \Omega/\kappa)},
  \end{split}\end{equation}
  
when we make the aforementioned identification $\Omega/\kappa=E+i/2$. Thus, for an accelerating observer, the Minkowski vacuum looks like a thermal distribution of particles  with temperature  given in terms of the acceleration
\beq
T=\frac{\kappa}{2 \pi}
\eeq  
We see that the form of the thermal distribution is the same as the transmission coefficient across the IHO obtained in Eq. \ref{Tval}, except for the difference of sign in the denominator. The thermal distribution for the Hawking-Unruh effect for fermions is also shown to be with a positive sign\cite{Alsing}. Here we have considered the scalar field for simplicity.
We trace back the sign difference in the scattering context to the factor of 1/2 in the IHO eigenmodes, that we highlighted before. (To be specific the thermal factor comes from the square of the Gamma function, which has different values for $|\Gamma(i\Omega)|^2 \propto \frac{1}{\sinh(\pi \Omega)}$ and $|\Gamma(i\Omega+1/2)|^2 \propto \frac{1}{\cosh(\pi \Omega)}$ ). 

Since the Minkowski vacuum appears thermal, we can write down a thermal density matrix associated with this vacuum. We have already established that the Rindler modes are identical to IHO eigenstates. Therefore, we employ the notation of incoming and outgoing basis of the IHO as defined in Sec. \ref{Stheory} . Operators related by a Bogoliubov transform will have vacua related by a squeezing transformation as shown below \cite{Stone13}:
\begin{equation}
\begin{split}
&\ket{0,in}=\\& N \text{exp} \bigg[ i \int_{-\infty}^{+\infty} e^{-E\pi}(\hat{b}_E^{out,+} \hat{b}_E^{out,-}+\hat{b}_{-E}^{out,-}\hat{b}_{-E}^{out,+}) \frac{dE}{2 \pi} \bigg] \ket{0,out}
\end{split}
\end{equation}

In the above we have introduced the $\hat{b}^{out,\pm}_E$ operator acting on the vacuum corresponding to the incoming basis $\ket{0,in}$ (outgoing basis $\ket{0,out}$ representing the positive or negative sides of the incoming (outgoing) axis as seen in Fig. \ref{fig:Shear}. A thermal density matrix can be obtained by tracing out states on the $-$side of the IHO barrier:
 \beq
 \hat{\rho} =\Sigma_i e^{-2\pi E_i} \ket{E_i ,+} \otimes \bra{E_i,+}.
 \eeq


To summarize,{we show that the emergence of thermal factors in the context of event horizons of black holes and the quantum Hall point contact set-up is rooted deeply in the structural parallel between the wavefunctions of the Rindler modes/boost eigen-modes and the IHO eigensystem. Thus, the  eigensystem of quantum mechanical modes in a relativistic setting is fully accessible in the quantum Hall system under point-contacts.} This is one of the central results of our work. 

 For completeness, we present an effective metric in the quantum Hall system following Ref. \onlinecite{Stone13}, and relate it to the discussion of Rindler spacetime in the previous section. The effective velocity of the electron in the quantum Hall system under the application of the potential $V(x,y)=\lambda xy$ is given $v_{eff,y}= \frac{\lambda}{eB}\partial V/\partial x = (\lambda e/B)y$. We see that a particle with a positive y-coordinate asymptotically moves along the positive y axis, and similarly for particles with a negative y-coordinate. No particle crosses over the origin. That point of no-return can be interpreted as the event horizon. An event horizon is a null-surface in spacetime where locally the light cones are always pointing inward, therefore allowing only  one-way transport. The existence of an event-horizon in a spacetime implies a very specific structure of the space-time metric (See Appendix for detailed discussion on structure of metric of spacetime with event horizons and the Rindler approximation). 
 The effective space-time metric with an `event horizon' in the quantum Hall problem was then given by M.Stone\cite{Stone13} 
 \beq
 ds^2 =-dt^2+\frac{1}{v_{eff}(y)^2} dy^2
 \eeq
  Using the $v_{eff}= -\kappa y$ with $\kappa= \lambda/eB$, we can write the metric in Euclidean time $\tau=it$ as 
  \beq
  ds^2=\Omega^2(y)(dy^2+\kappa^2 y^2 d\tau^2)
  \eeq
  where $\Omega(y)=1/(\kappa^2 y^2)$ is a conformal factor. This is equivalent to the Rindler metric we had studied in the previous section. This map is "suggestive" of the Rindler metric in this non-Lorentzian context as pointed out in \cite{Stone13}, but it must be noted that the conformal factor blows up at the point of interest $y=0$.

The derivation of Hawking-Unruh effect presented above is one of many distinct ways of understanding this phenomenon \cite{UnruhWeiss,Witten,Robinson05,Parikh00,Bisognano,Sewell82,Vanzo11} and is the simplest one. The modern day understanding of the Hawking-Unruh effect and black hole thermality is in terms of entanglement \cite{Witten}. The key fact in that understanding is that the entanglement Hamiltonian is related to the Rindler Hamiltonian and thus to the boost generator/IHO. In the next section, we discuss the relation of IHO to the boost generator through Lie-algebra isomorphisms.



 \subsection{Massless Dirac equation on Rindler wedge and relation to IHO.}

The Rindler Hamiltonian is not an esoteric object from the view point of condensed matter physics. It is not restricted to description of quantum modes near an event horizon in spacetimes. The Rindler Hamiltonian has appeared in condensed matter setting in contexts as important as bulk-boundary correspondence and entanglement spectrum of topological phases \cite{Swingle12,Hughes16}. While a detailed discussion of those topics are beyond the scope of this paper, in this subsection we show that the IHO, the key object of our work IHO is related to  the zero-energy Dirac equation in the Rindler wedge in the zero mass limit. This specific Dirac equation is use to computed the entanglement spectrum and edge states of a quantum Hall system \cite{Swingle12,Hughes16}.

 The Dirac equation for zero-energy is given in 1+1 dimensions is by
 \beq
 (\slashed{\partial}+im)\Psi=0,
 \eeq
where $\slashed{\partial}=\gamma^\mu \partial_\mu$ and m is the mass. The Dirac spinor is given by $\Psi=(\psi_-, \psi_+)$. The Dirac matrices in 2 dimensions are given by $\gamma^0=\sigma^x, \gamma^1=-i\sigma^y$.
Now, we want to express this in Rindler co-ordinates $(\rho, \tau)$. Let us recall that in terms of the `light-cone' basis $(v,u)=(t+x,t-x)$, the transformation to Rindler wedge is given by $u=\rho e^{\tau},v=\rho e^{-\tau}$. Writing the left/right chiral modes $\psi_{\pm}$ as $\psi_{\pm}= e^{\pm \tau} \chi_{\pm}$, one obtains the following equation for the chiral fields for zero mass \cite{Sierra14}:

\beq
\pm i\bigg(\rho \partial \rho +\frac{1}{2}  \bigg)\chi_{\pm} = i\partial_{\tau} \chi_{\pm}
\eeq 

the Dirac equation for massless chiral fermions in the Rindler wedge reduces to the dilatation operator, which in turn can be canonically transformed into an IHO Hamiltonian.

The relation between the  the Dirac equation in the Rindler space and the dilatation operator has been recently  studied in its relation to the Berry-Keating Hamiltonian and the zeros of the Riemann zeta function \cite{Sierra}.  It was also studied in the context of realisation of the Rindler Hamiltonian and the Unruh effect in cold atomic lattice systems \cite{Laguna17}
 
In this section, we have demonstrated a parallel in the structure of quantum mechanical modes near an event horizon and the eigensystem of IHO, which manifests in the same thermal form emerging in both cases. The Hawking-Unruh effect is closely associated with scattering across a IHO barrier. We saw that the relation between IHO and the Lorentz boost plays crucial role in this discussion. In the next section we show that symmetry considerations suggest a deeper relation between the boost and IHO.


\section{Symmetry considerations}\label{Wigner}

\subsection{Symmetry parallels through isomorphisms}
 \begin{table*}
\begin{tabular}{|p{3cm}|p{6cm}|p{6cm}| }
\hline
&  Hawking-Unruh   &  Lowest Landau Level \\
\hline 
Platform  & Spacetime $(x,t)$   &   Non-commutative plane $[R_x,R_y]=-i\ell _B^2$ in LLL \\ 
\hline
Invariant structure  &  Spacetime metric $ds^2=dt^2-d\vec{x}^2$ &  Commutation relation $[R_i,R_j]=-i \ell_B^2 \epsilon_{ij}$  \\ \hline
Symmetry transformations  & Metric preserving Lorentz transformations  $\mathfrak{so(2,1)}$:
Boost $(K_1,K_2)$ and rotations $K_0$  & Area preserving potentials $\mathfrak{sl}(2, \mathbb{R})$: Shears/Saddle $(K_1,K_2)$ and rotation/Harmonic $K_0$ \\
\hline
Rindler Hamiltonian & Boost & Inverted Harmonic oscillator\\
\hline
\hline

\multicolumn{3}{| c |}{Algebra of transformations   $[K_1,K_2]=-iK_0$ , $[K_0,K_1]=iK_2$ ,  $[K_2,K_0]=iK_1$}    \\
\hline
\multicolumn{3}{| c |}{$\mathfrak{so(2,1)} \approx \mathfrak{sl}(2, \mathbb{R})$} \\
\hline
\end{tabular}
\caption{Table highlighting the parallels between the symmetry structures and platforms in the Hawking-Unruh effect and the lowest Landau level}
\label{Table}
\end{table*}

The identity in the structure of the eigensystems of the Lorentz boost and the IHO realised in the quantum Hall setting bridges two very disparate physical settings. We point out that the parallel between the two structures can be put in a broader context of symmetry parallels between these different platforms. By this we mean an equivalence between the two set of objects- i) A class of potentials applied on a quantum Hall system ii) The group of Lorentz transformations acting on a space-time. Each of these platforms have an invariant structure and there are symmetry transformations that preserve this invariant. We point out that there is an isomorphism between the Lie-Algebras of the generators of the symmetry transformation in each case. The Lie-algebra isomorphisms are only suggestive of an equivalence between the generators and their action on a Hilbert space of states as a Hamiltonian. But an exact equivalence requires a much rigorous mathematical proof, which is beyond the scope of this paper.

 We start with the quantum Hall setting by recalling  the forms of the LLL potentials  identifying $P= R_x/\ell_B$ , $X= R_y/\ell_B$ and renaming $V_1,V_2,V_3$ as $K_0,K_1,K_3$ respectively. Then, from Eq.\ref{Eq.LLLpots} we have the Hamiltonians in the LLL generated by applied potentials to be of the form 
\beq
K_0= (P^2 +X^2), \quad  K_1 =(PX+XP), \quad K_2 =(P^2 -X^2)
\eeq
In this basis of LLL wavefunctions, these can be written as differential operators 
\begin{equation}\begin{split}
K_0= \frac{1}{4} \bigg(-\frac{\partial^2}{\partial X^2}+ X^2 \bigg)&,  \quad K_1 = \frac{i}{2}\bigg(X\frac{\partial}{\partial X}+\frac{1}{2} \bigg),\\ \quad K_2=  \frac{1}{4}&\bigg(-\frac{\partial^2}{\partial X^2}- X^2 \bigg)
\end{split}\end{equation}
These are exactly the generators of the Lie-algebra $\mathfrak{sl}(2, \mathbb{R})$ (in a given representation) \cite{ReadVisc}. The group $SL(2,R)$ consists to 2 dimensional matrices of unit determinant. One could think of these as area preserving deformations in two dimensions and there are three generators of such a deformation. To get some intuition, one could think of a square as shown in Fig.\ref{fig:Shear}: it can be rotated within its plane and the area does not change- this is done by the rotation generator $K_0$. One can stretch it sideways increasing the length and decreasing the width  preserving the area or one could deform it to a parallelogram. These two are the shear transformations. The non-trivial part is that the order of successive transformations do not commute, but the non-commuting part  will always be related to one of the generators. 
This is expressed as the $\mathfrak{sl}(2, \mathbb{R})$ `Lie-algebra':
\beq
[K_1,K_2]=-iK_0 ,\quad  [K_0,K_1]=iK_2, \quad   [K_2,K_0]=iK_1
\label{Eq.Isomorph}
\eeq
These generators are now the Hamiltonians acting on the LLL states.  One can also think of them as the generators of canonical transformations that preserve the commutation relation $[R_x,R_y]=-i\ell_B^2$.

 Now moving to the context of a (2+1)-dimensional Minkowski spacetime, there are generators of transformations of the spacetime $(t, \vec{x})$ such that the metric $ds^2=dt^2 -d \vec{x}^2$  is preserved. These generators form the Lie algebra $\mathfrak{so}(2,1)$  corresponding to the Lorentz group $SO(2,1)$.  There are three generators corresponding to one rotation and two boosts. The Lorentz group plays a fundamental role in quantum field theory in generating representation of one-particle states of fields \cite{Weinberg}.  These are the Lie-algebras corresponding to the distinct physical settings we are considering in this work.
 
 The exact mathematical relation between the two Lie algebras is a Lie algebra isomorphism\cite{Kim90}. 
 \beq
 \mathfrak{sl}(2,\mathbb{R}) \sim \mathfrak{so}(2,1)
 \eeq
obtained by identifying the Lorentz generators $(K_0,K_1,K_2)$ with the shear generators $(L,J_a,J_b)$ of Eqs.~(\ref{eq:j1}--\ref{eq:j3}). One can map from one setting to the other such that the algebraic relation in Eq.(\ref{Eq.Isomorph}) is always preserved.  {In this sense}, the shear generators or the saddle- electrostatic potentials are equivalent to the boosts and the rotation generators in LLL to rotation generators in spacetime in terms of the transformations they generate in the respective platforms. 
 One must be cautioned that this parallel at least at the level of Lie algebra isomorphism should not be taken too literally. We are not interested in treating the quantum Hall system as an effective spacetime as done in the field of analog gravity\cite{Barcelo}. We are particularly interested in the action of the generators on the quantum mechanical states. In the quantum Hall system, these generators are the Hamiltonians in the LLL and generate the time evolution of states.  A summary of the parallels in the structures between the two setting is given in the table \ref{Table}.
 

 From the above isomorphism, one immediately sees that the  two shear generators
 \beq
 K_2 =\frac{1}{4}(P^2 -X^2) \quad K_1= i\frac{1}{2}(XP+PX),
\eeq 
 are equivalent to the two boosts. These are just the Hamiltonians for the IHO in the position and the light-cone basis. Thus we have shown the equivalence between Lorentz boost and the IHO at the levels of wavefunctions, dynamics and operators. This provides a deeper analogy between the Rindler Hamiltonian dynamics and quantum mechanical barrier scattering. While the quantum Hall effect is the platform we consider in this work for the realisation of the IHO, there are many other viable settings such as quantum optics and  Josephson junctions.

 The use of Lie-algebra isomorphisms to relate two different physical scenarios has been done in a few earlier works. We have shown the equivalence between boost and IHO using Lie-algebras in 2+1 dimensions. The identification of the boost generator with the IHO is valid in 3+1-dimensions  as well as shown recently \cite{Betzios16}. 
 A similar Lie algebra isomorphism can also be used to realize Lorentz kinematics in quantum optics as well\cite{Kim90}, as discussed in Sec.~\ref{IHOforms}.

  The discussion in terms of these symmetry parallels was geared towards understanding appearance of Hawking-Unruh thermality in quantum Hall effect. But this analysis gives us much more. Now, we turn to other benefits of the structure laid out in our pursuit of a holistic understanding of the Hawking-Unruh effect in the LLL setting. now Along these lines, we explore the phenomenon of Wigner rotation, which is a manifestation of the non-commutation of boosts as given by the Lie-algebra.

 \subsection{Wigner rotation}
 
It has long been shown that two non-collinear Lorentz boosts give rise to a net boost in a given direction followed by a rotation. Mathematically, this statement is given as - 

\begin{align}
    S(\theta,\lambda)S(0,\eta)=S(\alpha,\xi)R(\phi)\label{wigrot}
\end{align}

Here, $S(\alpha,\beta)$ indicates a Lorentz boost at an angle $\alpha$ with respect to the $x-$axis, with boost parameter $\beta$ and $R(\phi)$ is a rotation by angle $\phi$. The angle $\phi$ is called the Thomas-Wigner angle. It is related to the parameters of the non-collinear boosts in the following way - 
\begin{align}
    \tan \frac{\phi}{2}=\frac{\sin\theta\tanh\frac{\lambda}{2}\tanh\frac{\eta}{2}}{1+\cos\theta\tanh\frac{\lambda}{2}\tanh\frac{\eta}{2}}\label{wigangle}
\end{align}
Similarly, it can also be determined that 
\begin{align}
    \cosh\xi&=\cosh\eta\cosh\lambda+\sinh\eta\sinh\lambda\cos\theta\\
    \tan\alpha&=\frac{\sin\theta\cos\theta(\sinh\lambda+\tanh\eta)(\cosh\lambda-1)}{\sinh\lambda\cos\theta+\tanh\eta(1+\cos^2\theta(\cosh\lambda-1))}
\end{align}

The Wigner rotation may also be cast in the language of squeezing transformations. The local isomorphism between the Lie algebras $\mathfrak{sp}(2,\mathbb{R})$, $\mathfrak{sl}(2,\mathbb{R})$, $\mathfrak{su}(1,1)$ and $\mathfrak{so}(2,1)$ have led to an interchangeable understanding of Lorentz boosts, and squeezing transformations. In terms of photonic or equivalently, harmonic oscillator states, Lorentz boosts correspond to area-preserving transformations in Wigner phase space, and hence, squeezing of the states. Quantum optics literature is replete with studies of squeezed states and their utility in interference experiments. By virtue of the Lie algebra isomorphism, these ideas maybe extended to quantum Hall systems. For instance, in Refs.~\cite{Smitha10,Varsha19} it was shown that the dynamics of anyons in quantum Hall point contact systems could be described by squeezed states.

Therefore, while the physics of squeezing in Wigner phase space and boosting in spacetime might be different, the equation \ref{wigrot} is fully applicable to both. Exploring Lorentz kinematics through quantum optics has actively been pursued through a variety of possible suggestions for experiments in quantum optics \cite{Yurke,Kim90,Ba_kal_2005,Baskal,Noz,CHIAO1989327}.

In essence, in these experiments one consecutively boosts  a coherent state twice, and compares its resultant parameters with those of a coherent state that has been rotated appropriately and squeezed. If the "same" state results from the two experiments, then the result serves as a confirmation of our expectations from Lorentz group kinematics. The central idea in these experiments boils down to the ability to accurately tune the parameters pertaining to squeezing of optical coherent states, or measuring the parameters of such squeezed states. From \ref{wigangle}, one might determine the Wigner angle. We propose that the application of the IHO onto the LLL, by means of strain operations or QPCs offers yet another domain for experiments on the measurement of Wigner rotation. Squeezed states are obtained in the LLL on time evolution by the IHO, and engineering Fabry-Perot or Mach-Zehnder type electron interferometry setups in the LLL \cite{Smitha10,Bocquillon,Varsha19} can determine the parameters associated with the squeezed state.

\section{Wave equation in black hole spacetime, IHO and  QNMs}
\label{WaveEq}


 In this last section, we study the physical manifestation of quasinormal modes in black hole physics. Quasinormal modes play an important role in black hole physics as their manifestly characteristic "excitations". Black holes are characterised only by their mass, charge and angular momentum. Due to their defining property of black holes as one-way membranes and the matter-energy trajectories are always directed towards the singularity. Thus, classically no information about the black hole properties is available for observers outside the horizon. But it was discovered in the pioneering work \cite{Vishveshwara} of C.V.Vishveshwara that through gravitational wave scattering one could obtain the parameters of a black hole from the decay rates and ringdown frequencies of quasi-normal modes. Since then QNMs have been object of intense study for multitude of physical considerations ranging from gravitational waves to holography \cite{Kokkotas,Konoplaya11,Berti}. QNMs have been one of the key measurements in the recent gravtiational wave detections through LIGO\cite{LIGO1}.

Here we describe the basic set-up for studying QNMs of black holes and its relation to the IHO resonances. We show that the IHO appears in a related scattering problem of an equation motion of a field in black hole spacetime under WKB approximation. We also clarify the distinction between the IHO appearing in this setting and that of Hawking-Unruh effect that we have already studied.

Wave equations  appear as equations of motion for fields that correspond to matter or energy on a background metric of a black hole. These could arise from perturbations of a scalar or higher spin fields on a background metric. One could also obtain the equation of motion for the perturbation of the metric in the Einstein equations, in which case the solutions correspond to gravitational waves. In each of these cases, under further simplifications the wave equations reduce to a `Schrodinger equation' for scattering against a potential.  The top of the scattering potential can be approximated as an IHO. The resonances thus obtained from the IHO are approximations to QNMs of the black hole spacetimes. As mentioned in Sec. \ref{Summary}, we again stress here that IHO appearing in this context is from a classical scattering scenario and we are not quantizing the fields under consideration.

 As the simplest instance, let us consider the equation of motion for a massless scalar field $\phi$ in curved space-time. This takes the Laplacian form \citep{Konoplaya11}
 \begin{equation}
 \Box \phi =\frac{1}{\sqrt{-g}}\frac{\partial}{\partial x^{\nu}} \bigg( g^{\mu \nu} \sqrt{-g} \partial_{\mu} \phi \bigg),
 \end{equation}
 where $g^{\mu\nu}$ is the metric of the space-time and $g$ is the determinant of the metric.For a Maxwell field (spin-1 gauge field),  Maxwell's equation in the absence of a source reads
\begin{equation}
\frac{1}{\sqrt{-g}}\frac{\partial}{\partial x^{\nu}} \bigg( g^{\rho \mu}g^{\sigma \nu} \sqrt{-g} F_{\rho \sigma} \bigg)=0
\end{equation}
 One can obtain similar equations of for higher-rank fields, such as perturbations of the metric itself. 
 
 
 In stationary and rotationally invariant spacetimes, we can always decompose the field into a ``radial part'' and an ``angular'' part. The angular part is given in terms of spherical Harmonics, $Y_{lm}$. For example, for a scalar field:
\begin{equation}
 \phi(t,r,\theta,\phi)=e^{-i \omega t} Y_{lm}(\theta, \phi) \psi(r)/r
 \end{equation}
Here $r$ is the radial co-ordinate for example in the Schwarzschild spacetime.  For the spherically symmetric Schwarzschild metric in the vicinity of a black hole, this substitution reduces the wave equation reduces to the following form in terms of an effective potential $V$
 \begin{equation}
\frac{\partial}{\partial r_*^2} \psi + V(r, \omega) \psi =\omega^2 \psi.
\end{equation} 
Here $r_*$ is called the `tortoise co-ordinates' defined as $r_*=r+2GM \log|\frac{r}{2GM}-1|$, and $M$ is the mass of the black hole.
This equation resembles a quantum mechanical problem of scattering off a potential barrier $V$. The effective potential $V(r,\omega)$ for a Schwarzschild spacetime is given by
\begin{equation}
V(r)= \bigg( 1- \frac{2M}{r}\bigg) (\frac{l(l+1)}{r^2} + \frac{2M(1-s^2)}{r^3})
\end{equation}
For scalar field the spin $s=0$, for Maxwell gauge fields $s=1$ and for gravitational perturbation of axial type $s=2$\cite{Schutz85,Chandrasekhar}. 

Let us perform a semiclassical analysis of our wavefunctions near the peak of the effective potential\citep{Schutz85}. Suppose we consider an incident wave that does not have enough energy to pass through the potential. Letting $r_1$ and $r_2$  denote turning points of the potential $V$, the WKB  wavefunction outside the turning points is given by \citep{Schutz85}
\begin{equation}
\psi_i (r) \approx V^{-1/4} exp \bigg( \pm i \int_r^{r_i} [V(r')]^{1/2} dr' \bigg),
\end{equation} 
where $i=1,2$. Within the region between the turning points, the potential can be approximated by a parabola. 

Thus, we can expand around the peak of the potential at $r=r_0$ to get
\begin{equation}
V(r)=V_0 +1/2 V_0 '' (r) (r-r_0)^2 + O((r-r_0)^3).
\end{equation}
Then the scattering equation becomes 
\begin{equation}
-\frac{d^2 \psi}{dx^2}+ +\frac{1}{4}x^2\psi =(\nu +\frac{1}{2}) \psi,
\end{equation}
where $x= (2V_0 '')^{1/4} e^{i \pi/4}(r-r_0)$, $\nu+1/2 =-i V_0 /(2V_0 '')^{1/2}$.
This is the Weber equation, which can also be seen above to take the form of Schrodinger's equation in the quantum mechanical problem of scattering against an IHO potential. The effective potential is parametrized only by the mass of the black hole. In the approximation that gives rise to IHO potential, the curvature of the effective potential is also a function of the mass of the black hole.
This derivation shows an equivalence between the classical scattering in a gravitational spacetime and a quantum mechanical scattering problem.

The general solution of the Weber equation/ IHO Schrodinger equation is given by
\begin{equation}
\psi = A D_\nu (x) +B D_{-\nu-1}(ix).
\end{equation}
The asymptotic forms at large $x \rightarrow \infty$ is given by \citep{Schutz85}
\begin{equation}
\begin{split}
\psi  &\sim B e^{-3i\pi \frac{\nu+1}{4}} (4k)^{-\frac{(\nu+1)}{4}}(x-x_0)^{-(\nu+1)}e^{i\sqrt{k}\frac{(x-x_0)^2}{2}}\\
&+[A+B \sqrt{2\pi}\frac{e^{-i\nu \frac{\pi}{2}}}{\Gamma(\nu+1)}]e^{i\pi \frac{\nu}{4}}(4k)^{\frac{\nu}{4}}(x-x_0)^\nu e^{-i\sqrt{k}\frac{(x-x_0)^2}{2}}
\end{split}
\end{equation}
  Here $k=1/2 Q_0 ''$ and $\Gamma(\nu)$ is the Gamma function. To obtain purely outgoing wave, set the co-efficient of the incoming wave to zero. This involves finding $\nu$ such that $\Gamma(-\nu)=\infty$. This leads to a condition that $\nu$ can take only integer values
  \begin{equation}
  Q_0/\sqrt{2 Q''_0}=i(n+1/2) \quad n=0,1,2...
  \end{equation}
  Therefore, by demanding asymptotically purely outgoing or incoming states, one obtains purely imaginary `eigenvalues' of the effective Schrodinger equation. These are the famous `quasi-normal modes' of black holes \cite{Vishveshwara}. Thus, the WKB approximation for scattering against the effective potential of the black hole spacetime leads to and IHO problem and the resonances correspond to the quasi-normal modes. 
  
  The effective potentials for fields on black hole spacetimes have additional features such as local minima, asymptotic behaviour that are beyond the WKB approximation. Upto sixth order corrections are computed over the WKB approximation \cite{Berti}. The WKB approximation of black hole QNMs with IHO resonances are suitable only for low $n$ and large angular momentum channels. The main crucial fact that the IHO approximation captures is that the  quantised decay rates of QNMs are functions of the curvature of the potential maximum which is dependant on the black hole mass. One could also approximate black hole QNMs with the resonances of P\"{o}schl-Teller potential and Eckart potential \cite{Mashhoon,Ferrari,Aguirregabiria}.

\section{Outlook}
\label{Outlook}
We have seen how the inverted harmonic oscillator serves as a central unifying model, tying together diverse physical systems such as point contacts in the quanutm Hall effect, scattering from black holes, the Hawking-Unruh effect, and Lorentz transformations. Below we provide a thorough survey of the manifestation of IHO in diverse areas of physics from string theory to chaos. This provides a road map for further exchange of ideas between different fields and possibly exploring them in the quantum Hall platform. Then we address another important topic, namely entanglement, and discuss how it relates to the topic of Hawking-Unruh effect and certain fundamental aspects of thermality and entanglement.

\subsection{Roadmap to topics and works associated with the IHO}\label{survey}

Let us now summarize--for the benefit of the reader--some of the diverse contexts in which the IHO has appeared. The IHO has been studied for its scattering properties in the texts of Kemble\cite{Kemble37} and Landau and Lifshitz \cite{Landau}. Its interesting semiclassical aspects were studied by Ford and Wheeler \cite{Ford59}, with subsequent applications to nuclear decay and emissions \cite{Ford59}. A thorough study of scattering amplitudes, wave packet dynamics, delay times, and the role of dissipation was done in the thesis of G.Barton \cite{Barton86}, which arguably remains to this day the most comprehensive analysis of the IHO. More modern and mathematical analyses of the IHO can be found in works of Chruscinski \cite{Chruscinski03,Chruscinski04},Yuce\cite{Yuce}, Shimbori and Kobayashi \cite{Shimbori00,Shimbori1,Shimbori}. Another important context is that of a parametrically driven SHO (where the frequency is varied periodically in time). The IHO appears in this setting in a particular limit\cite{Perelomov}. The IHO has appeared in various physical contexts, such as thermal activation \cite{Boyanovsky} and cosmological inflation \cite{Guth}. Its role in instability, dissipation, and decoherence of open systems have been studied using von Neumann entropy, Loschmidt echoes and Wigner function analysis \cite{Miller,Cuchietti,Zurek94}. The IHO has been known in the quantum optics community as the ``Glauber oscillator,'' and has been experimentally realised \cite{Gentilini}. It also plays a role in generating squeezed states of light, which have been extensively investigated both theoretically and experimentally in the quantum optics literature\cite{Yurke,Lvovsky,Teich,Sudarshan,Wu}. There are some intriguing paradoxical features in construction of Wigner function to describe quantum mechanical tunneling across an IHO in phase space. \cite{Balazs90,Heim13}. The IHO has been studied in the context of string theory as well as the $c=1$ matrix model \cite{Cremonini,Friess,Maldacena05,Schoutens,Dhar93,Dhar92,Karczmarek04}. The IHO is related to the `Berry-Keating Hamiltonian', which has been proposed to count the zeros of the Riemann zeta-function and has signatures of quantum chaos \cite{Berry,Sierra,Bhaduri,Bhaduri97}. More recently, the IHO has caught attention again in the context of chaos and Lyapunov bound \cite{Morita19,Morita18}. This is closely related to the appearance of chaos in black hole dynamics \cite{Hashimoto17,Dalui19,Dalui192}. The quasinormal modes of black holes are known to be captured by the resonances of the IHO \cite{Schutz85,Mashhoon,Ferrari}. A much deeper connection between the Rindler Hamiltonian capturing near-horizon quantum mechanical behaviour and the IHO has also been shown using projective light-cone construction methods \cite{Betzios16}. 

Moving from fundamental quantum mechanical aspects to many-body, condensed matter settings, the IHO has also made appearance since the early days of electron theory of metals. The IHO appeared there in the context of magnetic breakdown of metals due to intra-band semi-classical tunneling of trajectories in electron bands in magnetic field \cite{Azbel}. A more recent work on modern semiclassical theory of magnetic breakdown makes central use of IHO scattering physics \cite{Alexandrinata} . Of direct relevance to the current work, the IHO has appeared in the quantum Hall systems under the application of a saddle potential as has been seen in early work of Fertig and Halperin \cite{Fertig87}, detailed above.

The realisation of the IHO in quantum Hall systems and its relation to the Lorentz boost and the Rindler Hamiltonian was pointed by the authors recently\cite{Hegde19}. Since then there have been several recent works which have highlighted the appearance of IHO in the context of horizon thermality, chaos and complexity \cite{Hashimoto20,Bhattacharyya19,Bhattacharyya20,Ali20,tian20,Betzios20, Dalui20,Dalui19,Dalui192}.

This concludes a brief survey of the quantum mechanics of the IHO covering many intriguing aspects. In the below, we discuss the entanglement aspect of Hawking-Unruh effect and show how the Rindler Hamiltonian appears as the entanglement Hamiltonian.

 \subsection{Half-space entanglement and thermal density matrix of the Rindler wedge}
  
Now, let us consider the entanglement properties of some degrees of freedom  defined in Minkwoski space-time (say some field  $\phi$ defined over a time-slice t=0). As we have explained in the previous sections, the Rindler observers and their corresponding spacetime are restricted by the lightcone structure. The right Rindler wedge is causally disconnected from the left Rindler wedge and provides a natural partition of the Minkowksi spacetime. Let us consider a state characterised by the density matrix $\rho$. The system is partitioned into two sub-regions : $x>0$ denoted by A and $x<0$ denoted by $\bar{A}$.The `causal development' of region A is the right Rindler wedge as shown in Fig.\ref{fig:Rindler}. We want to calculate the reduced density matrix $\rho_A$ corresponding to the region A, by tracing/integrating out the degrees of freedom in regions $\bar{A}$:
\beq
\rho_A = \text{Tr}_{\bar{A}}\rho
\eeq 

The entanglement Hamiltonian or the `modular Hamiltonian' is defined as the logarithm of the reduced density matrix
\beq
H = -log(\rho_A).
\eeq
The spectrum of this Hamiltonian is called the entanglement spectrum and is one of the powerful tools in characterising the entanglement properties of the system and it encodes information on all the Renyi entropies. 

 Usually the entanglement Hamiltonian is a highly non-local object and does not allow for a neat analytical expression. Among few instances where it does take a local form, the case of the Rindler wedge is one. For the partition of the Minkwoski space into the Rindler wedges, the Entanglement Hamiltonian of the reduced density matrix for the right Rindler wedge is given by  
\beq
\rho_A =\frac{e^{-2\pi H_R}}{Tr(e^{-2\pi H_R})},
\eeq
where $H_R$ is the Rindler Hamiltonian \cite{Witten}.  The above expression is exactly the form of a thermal density matrix with temperature $T=1/2\pi$. The crucial point here is that the Rindler Hamiltonian is given by the boost generator restricted to the right wedge. 

We give a bird's eye view into the derivation of the above equation. A full derivation with all the subtleties can be found in Refs.\onlinecite{Witten,Padmanabhan19}. Consider a configuration of scalar field $\phi$ in Minkwoski spacetime. The vacuum wave functional starting in some arbitrary state $\ket{\chi}$ and evolves to the state $\ket{\phi}$ , can then be written as a Feynman path integral over different field configurations
\beq
\Psi_0[\phi] = \lim_{t \rightarrow \infty} \bra{\phi} e^{-itH} \ket{\chi} =\int_{\phi (\tau \rightarrow -\infty)=\chi}^{\phi(t=0)=\phi} \mathcal{D}\phi e^{-S/\hbar}
\eeq
Here $S$ is the Euclidean action corresponding to the Hamiltonian $H$. The initial state is given on a $t=0$ hypersurface on the whole range of $-\infty < x < \infty$. Consider a cut that separates Minkowski spacetime into two regions belonging to left and right Rindler wedges. The field configurations in those regions is indicated by $(\ket{\phi_R},x>0),( \ket{\phi_L} x<0)$. 


The time-like co-ordinate in the right Rindler wedge $\tau$ can be used as an angular co-ordinate with periodicity $2\pi $ as explained in the previous section. The time evolution in Minkowski time $t =-\infty$ to $t=0$ is therefore mapped to evolution along the angular co-ordinate from $t=\pi \rightarrow 0$. One can make use of this mapping between the two to slice the path-integral in two different ways.That is, the lower half-plane of the path integral is covered in two different ways for the Minkowski and Rindler time co-ordinates co-ordinates. On one hand one can slice the path integral into integration over field configurations at constant time slices. At each time slice the range of $-\infty < x < \infty$ is covered in the Minkowski spacetime. On the other hand, for $\tau=0$, the range of $\infty> x>0$ is covered in the Rindler co-ordinates. Then the path integral becomes
\beq
\Psi_0[\phi] = \int_{\phi (\tau \rightarrow \pi/\kappa)=\phi_L}^{\phi(\tau=0)=\phi_R} \mathcal{D}\phi e^{-S/\hbar}
\eeq
 Due to the partition from the Rindler wedge the state $\bra{\phi}$ is  now a tensor product of $\bra{\phi_L} \otimes \bra{\phi_R}$. The rotation in the angular Rindler time is generated by the Rindler Hamiltonian $H_R$. Therefore, the path integral takes the form
 \beq
 \Psi_0[\phi_L ,\phi_R] = \bra{\phi_R} e^{-\pi H_R } J \ket {\phi_L}
 \eeq
The final state $\phi_L$ is written as  $J\ket{\phi_L}$, where $J$ is the PCT(parity, charge conjugation and time-reversal) conjugation operator which comes along with the Lorentz symmetry

 Now integrating over the degrees of freedom on the left one obtains the 
  \beq
  \rho = \text{Tr}_L \ket{0} \bra{0} = e^{-2\pi H_R}
  \eeq
    The above discussion  shows that the quantum mechanical degrees of freedom on the right and left Rindler wedge are entangled and restricting to one of the wedge `integrates' over the other to give a thermal density matrix. We see that the entanglement Hamiltonian from the above reduced density matrix is actually the Rindler Hamiltonian. 
    
This is an important and subtle point to be appreciated: 
 -{What was considered an entanglement Hamiltonian for a hypothetical entanglement cut of the space-time has a manifestation as a physical Hamiltonian that generates dynamics in a physical sub-region of the spacetime that is the Rindler wedge. The reduced density matrix for that entanglement cut is now a physical thermal density matrix.} The form of the this Hamiltonian is exactly known. The Lorentz boost restricted to the right Rindler wedge is the Rindler Hamiltonian. The detailed derivation of the Hawking-Unruh effect was restricted to non-interacting field, nevertheless showed the important role played by the boost generator. The path-integral derivation is much more general in assumptions about the quantum field considered and shows a much deeper connection between entanglement and Hawking-Unruh thermality  making manifest the identity between entanglement Hamiltonian and Rindler Hamiltonian. The parallel we have shown between the boost eigensystem and that of IHO thus opens up the possibility to probe entanglement aspects of the phenomenon in future. 

Before ending this section we note that the entanglement Hamiltonian is in general a highly non-local quantity and seldom can be written in terms of local operators. The fact that it is possible in the context of Rindler wedge is consequence of a fundamental theorem called the Bisognano-Wichmann theorem, the statement of which can be summarised that the vacuum is a thermal state(as defined by Kubo-Martin-Schwinger condition) with respect to the generator of Lorentz boosts\cite{Bisognano}.
The machinery underlying this theorem has played a key role in the field of quantum gravity in defining constraints on energy content in a given region of spacetime \cite{Witten,Jefferson,Balakrishnan,Lashkari}. Even more importantly, the underlying machinery lie at the foundations of quantum statistical mechanics\cite{Haag,Bratteli} in deriving Gibbs ensemble starting from operator algebras of quantum mechanics. While we are not directly engaging with these foundational topics in this work, we mention them for the interest of a curious reader interested in broader foundations of physics.

To relate the above discussion back to IHO in quantum Hall systems, we saw that the IHO is closely related to the Lorentz boost generator. We also saw that there is a Lie-algebra isomorphism between the generator of Lorentz group $\mathfrak{so}(2,1)$ and area preserving deformations/ quantum Hall potentials $\mathfrak{sl}(2,R)$. It would be interesting to make this relation more precise by capturing all the structures such as the Hilbert space and its representation, the action of the the generators of Lie-algebra on the Hilbert space and the path-integral in the context of $\mathfrak{sl}(2,R)$ and relate to  quantum-Hall observables. This would entanglement studies in quantum point contact systems into a different perspective, given that point contacts have already been studied in the light of full counting statistics and entanglement.

\subsection{Conclusion and future directions}

As mentioned previously, this work is equal parts new results and review. We showed in a gauge-invariant way how a variety of physically relevant quadratic potentials could lead to IHO dynamics in the lowest Landau level. After reviewing the physics of the IHO and the formalism of time-dependent scattering, we showed how time-resolved scattering measurements of electrons in a quantum Hall system can reveal the signatures of quantised time-decay of QNMs. Even though the IHO is a rather idealized model for a quantum point contact, we showed that the qualitative features of these results are universal, by studying the more realistic P\"{o}schl-Teller potential. We showed how the IHO appears in the context of black hole physics in the guise of the Rindler Hamiltonian, where the same transformation of states that led to the S-matrix in quantum Hall scattering here describes the spectrum of Hawking-Unruh radiation in an accelerated vacuum. Finally, we closed the loop by showing the explicit algebraic mapping connected these diverse concepts.

 Going forward, our results can be extended in several directions. First, our results on time-dependent scattering in quantum point contacts can be applied to a diverse array of experimentally-relevant mesoscopic devices. Preliminary analysis on graphene-based quantum Hall devices\cite{Hegde19} shows that the quasinormal mode frequencies for point contact potentials may be measurable in experiment in the near future. Additionally, the algebraic parallels between the IHO and Lorentz boosts opens the door towards simulation of (special-)relativistic effects in quantum Hall systems. As we showed, a quasiparticle in the lowest Landau level passing through a point contact potential undergoes a distortion that is algebraically equivalent to a Lorentz boost. By arranging different configurations of point contact potentials, phenomena such as the non-commutation of Lorentz boosts (Wigner rotation) can be probed experimentally. As the phase due to a Wigner rotation is intimately related to the Hall viscosity\cite{ReadVisc,Bradlyn2012,haldane2009hall}, this offers a route towards an indirect measurement of the nondissipative viscosity. From a more theoretical standpoint, it would be interesting to generalize our work to correlated states of matter such as the fractional quantum Hall (FQH) effect. We expect that (anyonic) quasihole states in FQH systems will transform under more exotic representations of the IHO Lie algebra, which could be probed in point contact and interferometer setups. This would build off recent work studying squeezing of anyonic quasiholes\cite{Varsha19}. Finally, our framework could be applied to the study of entanglement in condensed matter systems in a more general fashion to draw more concrete parallels to questions in black hole thermalization. To conclude,  by placing the inverted harmonic oscillator as the central concept, we have offered common ground for linking a  web of conceptual connections between quantum condensed matter physics and relativity.


\begin{acknowledgements}
 We are fondly indebted to the guidance of late black hole physicist, mentor, and father, C. V. Vishveshwara, whose pioneering prediction of quasinormal modes in black holes was marked by its $50^{th}$ anniversary this year. We sincerely thank Vatsal Dwivedi, George Fuller, and Michael Stone for illuminating discussions. We acknowledge the support of the National Science Foundation under grants DMR-1945058 (B. B.) and  DMR-2004825 (V. S.).  B.~B.~acknowledges the support of the Alfred P. Sloan foundation. S.~V.  is very grateful to UC San  Diego's Margaret Burbidge Visiting Professorship Award and the close-up discovery it brought of Burbidge as an incredible inspiration. 

\end{acknowledgements}

\section*{Appendix A : Rindler spacetime and Rindler approximation for black hole spacetimes.}
\label{Appendix}

\subsection*{The Rindler Wedge}
   \subsubsection{Minkowski space-time and Boosts}
   
   The Minkowski spacetime is described by its metric expressed below in  co-ordinates $(t,\vec{x})$
   \beq
   ds^2 = dt^2 - d\vec{x}^2 =\eta_{\mu \nu} dx^\mu dx^\nu,
   \eeq
   where $t$ is the time-like co-ordinate and the $x^i$ are space-like co-ordinates. {We shall set the velocity of light $c=1$.} Here we will restrict the discussion to 2+1 dimensional space-time for the convenience of later purposes, though all the following derivations are done in 3+1- dimensions.
  
  The space-time manifold has a `light-cone' structure defined by $ds^2=0$ as shown in Fig.\ref{fig:Rindler}. The light-cone underlies the causal structure of the spacetime in that it sorts the regions into time-like, space-like and light-like connected regions. In this sense, it is a `horizon' for observers at every point in spacetime. This aspect becomes important later.
  
  The metric and thus the lightcone is preserved by a set of transformations  involving translations, rotations, Lorentz transformations (from here on referred to as Boosts). This forms the Poincare group. The rotations and boosts form a subgroup called the Lorentz group. We will be particularly interested in the Lie algebra of the generators of this group. This algebra is represented as $\mathfrak{so}(2,1)$. This group comprises of a generator of rotation and two generators of boosts differing in direction by $\pi/4$.
  
  To recall, the boosts involve transformations of both space and time co-ordinates equivalent to moving to a reference frame moving with a velocity parameter $\beta$: $x'=\gamma(x-\beta t)$, $t'=\gamma(t-\beta x)$, where $\gamma=1/\sqrt{1-\beta^2}$. The inherent underlying hyperbolic nature of the Lorentz group is transparent when boosts are expressed in terms of hyperbolic transformations involving the 'rapidity' parameter $\theta= \tanh^{-1} \beta$: $x'= x\cosh \theta -t \sinh \theta$, $t'= t \cosh \theta - x \sinh \theta$. Let us note here that in going to Euclidean space-time $(x,\tau)$, $\tau =it$ , the boost is a rotation in $(x,\tau)$ with imaginary rapidity angle $i \theta$.
  
  Finally, let us also define the 'light-cone' co-ordinates: $u=t-x$, $v=t+x$. This makes boosts take a particularly simple form. On a boost with rapidity $\theta$, the lightcone co-ordinates transform as $(u,v) \rightarrow (u e^\theta, v e^{-\theta})$.

   \subsubsection{Uniformly accelerated observers}
   We will be particularly interested in the restricted region of the spacetime which is the spacelike regions subtended by the lightcones, which is called the Rindler wedges. Often, one is interested only in the left Rindler wedge. This is motivated by the physical situation of uniformly accelerating observers whose trajectories are restricted to these wedges. In the context of black holes, the observers outside the event horizon are also restricted to the region (conformally) equivalent to the wedge region.  
   
   We want to calculate the trajectory $(x(\tau), t(\tau))$ of the uniformly accelerating observers with acceleration $a$. The relativistic velocity parametrized by proper time $\tau$ along the observer's trajectory  is $u^\mu(\tau) = \frac{dx^\mu}{d \tau}= (\gamma ,\gamma \vec{v})$, where $\gamma$ is the Lorentz factor $1/ \sqrt{1-v^2}$ and $\vec{v}$ is the usual velocity. Using the condition $\eta_{\mu \nu} u^\mu u^\nu =1$ and the acceleration defined as  $a^\mu (\tau)= \dot{u}^{\mu}(\tau)$ is orthogonal to the velocity $\eta_{\mu \nu} a^\mu u^\nu=0$ . This leads to a covariant condition for constant acceleration \cite{LandauCTF}
\beq
\eta _{\mu \nu} a^{\mu} a^{\nu} =-a^2
\eeq .
 From this one obtains the equation $\frac{d}{dt}(v/\sqrt{1-v^2})=a$. From this the trajectory of a uniformly accelerating observer is given by a branch of the hyperbola $x^2 -t^2 =a^{-2}$.
 
 Now, to get the trajectories $(x(\tau), t(\tau))$, remember that the proper time $\tau$ is related to the Minkowski time $t$ by : $\tau= \int_0^t dt' \sqrt{1-v(t')^2}$. From this the trajectories are obtained:
 
\beq
t(\tau)= \frac{1}{a} \sinh(a\tau) \quad x(\tau)= \frac{1}{a} \cosh(a \tau)
\label{Eq.Traject}
\eeq

   \subsubsection{Rindler spacetime - The Wedges}

{\it Rindler Spacetime and Rindler horizon:  } The trajectories derived for a uniformly accelerating observer by themselves form geodesics of a spacetime called the `Rindler spacetime'.   Rather than considering trajectories of observers in Minkowski spacetime, we can study the Rindler spacetime and its causal structure. They are solutions of the Einstein's equations. This spacetime plays a very important prototype for understanding the black hole thermal physics.

The trajectories Eq.(\ref{Eq.Traject}) with positive acceleration are restricted to the `right wedge' of the Minkowski spacetime: $x>0$ and $|t|<x$.  The lightcone $t-x=0$ acts as a `horizon' for these set of observers: The observers in that part of the spacetime cannot access any information on $t>x$. Similarly, the accelerating observers with negative $a$ are restricted to the `left wedge' and have a horizon. We are primarily interested in such horizons in spacetimes, which when considered along with quantum mechanics give interesting results. An observer at rest in the Minkowski spacetime$(T(t),X(t)=(t,x))$ can receive information from the region $t>x$ and do not perceive such a Horizon. Therefore, this notion of horizon is an observer dependent concept with a dependence on the family of causal curves we consider.

 A broad class of spacetimes including curved spacetimes of gravity, with a metric that is `conformally' equivalent to the Minkowski metric are relevant for studying the horizon physics. Such a metric has the form:
\beq
 ds^2 =\Omega (\xi^0,\xi^1) ((d\xi^0)^2-(d\xi^1)^2),
 \eeq
where the Conformal factor $\Omega(\xi^0,\xi^1)$ is a non-zero function. The trajectories of the light rays form the `light-cone'.
 If one has a family of curves such as Eq.(\ref{Eq.Traject}) then $t-x$ act as a horizon. For a co-moving observer with proper time $\tau$, $(\xi^0(\tau),\xi^1(\tau))=(\tau, 0)$ on such a trajectory.  The change of co-ordinates from the Minkowski spacetime to a conformally equivalent spacetime is then determined from that condition to be-
\beq
\xi^0(\tau)= \frac{e^{\kappa \xi^1}}{\kappa} \sinh(\kappa \xi^0) \quad \xi^1(\tau)= \frac{e^{\kappa \xi^1}}{\kappa} \cosh(\kappa \xi^0)
\eeq
Here the acceleration of the observers is replaced by the parameter $\kappa$ of the Rindler spacetime and we shall stick to this use.
The metric now reads:
\beq
ds^2 =e^{2\kappa \xi^1} ((d\xi^0)^2-(d\xi^1)^2).
\eeq
These co-ordinates have a range: $- \infty < \xi^0 < =\infty$,  $-\infty < \xi^1 < + \infty$. This covers only the quarter of the Minkowski spacetime i.e the `right Rindler wedge'. The family of $\xi^0= \text{constant}, \xi^1 = \text{constant}$ curves are shown in Fig.\ref{fig:Rindler}.
The family of accelerated observers also cannot perceive distances larger than $\kappa^-1$ in the direction opposite to the acceleration  \cite{Mukhanov}. An observer at a space-like co-ordinate $\xi^1=0, \xi^0=0$ (an observer with acceleration $\kappa$ in Minkowski basically) measures an infinite range of co-ordinates $-\infty < \xi^1< 0$, measure the distance:
\beq
d=\int_{-\infty}^0 \exp{\kappa \xi^1} d\xi^1 =\frac{1}{\kappa}
\eeq
This is restating that an accelerating observer cannot measure the entire Minkowski spacetime and is bound by the horizon.  
   One can change to a different space co-ordinate $(1-\kappa \bar{\xi}^1)=e^{\kappa \xi^1}$ to write the metric as
   \beq
   ds^2=(1+\kappa \bar{\xi}^1)(d\xi^0)^2- d(\bar{\xi}^1)^2 
   \eeq
   Finally, making a transformation $(1+\kappa \xi^1)=\sqrt{2\kappa \ell}$ and $\tau=\xi^0$ we get,
\beq
 ds^2 = (2 \kappa \ell)d\tau^2- \frac{d \ell^2}{2\kappa \ell} 
 \eeq
 This form of the metric is quite  important  especially for focusing on the physics near a horizon. The horizon in this case is located at $\ell=0$. The advantage of this metric is that it can be extended to negative values of  $\ell$ to cover all four quadrants of the spacetime \cite{Padmanabhan19}.

  As will be shown below the Rindler metric is extremely  important for the following reasons: 
 \begin{itemize}
 \item  Other curved spacetimes with horizons can be approximated to the Rindler spacetime of the above form near their horizon.
 \item  This set of co-ordinates are also useful in making an `extension' to cover the full Minkowski spacetime. This is technically known as the `Maximal extension' and is used in extending the Schwarzschild co-ordinate that covers spacetime outside the black hole to Kruskal-Szekeres co-ordinate that covers the entire spacetime except the singularity.
 \item A non-trivial transformation between how the clocks tick (time translation) in the Rindler space and the completed space is the key feature in giving rise to thermality near horizons and black holes. 
 \end{itemize}

   \subsubsection{Black holes}
 The simplest definition of a black hole is that it is a one-way membrane \cite{CVVOneway}. Causal trajectories of matter and energy can only traverse through them in one-direction. A black hole is a `null-surface' or loosely speaking a trapped `light-wave' in that locally the normals to the surface are light-like. Thus, they are `horizons' beyond which information cannot be accessed by observer lying outside the black holes. Black hole spacetimes such as Schwarzschild, Kerr, and Reissner-Nordstrom are also exact solutions to Einsteins equations. Another simplest instance where one can see this horizon behavior is in the context of a Minkowski spacetime, which is flat and endowed with a lightcone structure.  This lightcone acts as a horizon for accelerating observers. In fact, the spacetime near the black hole horizons can be approximated in terms of `Rindler spacetime'\cite{Rindler}, which is associated with the description of spacetime for accelerating observers in Minkowski spacetime.

 \subsubsection{Rindler approximation to black hole horizons}
 Now we shall show that a general curved static spacetime with a horizon can be approximated near its Horizon can be to Rindler spacetime. This discussion will closely  follow the elegant line of reasoning similar to \cite{Padmanabhan19}. We saw that the Schwarzschild metric is of the form:
 \beq
 ds^2 =f(r)dt^2 -\frac{dr^2}{f(r)}+r^2 d\Omega^2, \quad f(r)=\bigg( 1-\frac{2M}{r}  \bigg)
 \eeq
 The horizon in this case was located at $r=2M$. One can in fact consider metrics of the above form with an arbitrary function $f(r)$, that has a simple zero at $r=r_H$, which is the event horizon. The near the horizon, the metric can be written as (focusing only on the time-like and space-like part of the metric) :
 \beq
 ds^2 \approx f'(r_H)(r-r_H)dt^2 - \frac{dr^2}{f'(r_H)(r-r_H)}
 \eeq
 Introducing $\kappa =f'(r_H)/2$ and $\ell =(r-r_H)$, the metric reads as 
 \beq
 ds^2 \approx 2\kappa \ell dt^2 -\frac{d \ell^2}{2\kappa \ell}
 \eeq
  This is exactly the form of the Rindler metric we derived in the previous sections. This derivation is made rigorous for a general spacetime in the following. This illustration that any black hole horizon spacetime approximates to Rindler spacetime near the horizon is extremely important especially for the Hawking-Unruh effect as the aspects of Rindler spacetime directly feed into its derivation. 
 
 Consider a spacetime $ds^2= g_{\mu \nu}dx^\mu dx^\nu$with the following conditions: 
 \begin{enumerate} 
 \item Static in that given co-ordinate representation of the metric $g_{0\nu}=0, g_{ab}(t,x)=g_{ab}(x)$
 \item $g_{00}(x)=N^2(x)$  that vanishes on a 2-hypersurface $\mathcal{H}$. The hypersurface is defined by the equation $N^2=0$.
 \item $\partial_\mu N$ is finite and non-zero on $\mathcal{H}$
 \item All other metric components remain finite and regular on $\mathcal{H}$
 \end {enumerate}
 The metric is then written as:
 \beq
 ds^2 =N^2(x^a )dt^2 -\gamma_{ab}(x^a)dx^a dx^b
 \eeq
 Now, a family of observers can be constructed similar to the accelerating observers in Minkowski spacetime. These observers are characterized by $\vec{x}=constant$, four-velocity $u_\mu=N\delta^0_\mu$ and four acceleration $a^\mu = u^\mu \partial_\mu u^\mu =(0,-a) $. The spatial components of this are given by $a_a=(\partial_a N)/N$. The unit normal to the hypersurface $N= constant$ is $n_a =\partial_\mu N (g^{\mu \nu} \partial_\mu N \partial _\nu N)^{1/2}= a_a(a_b a^b)^{-1/2}$. The normal component of the four-acceleration is related to the surface gravity $\kappa$.

 We can go to a co-ordinate where N is treated as one of the spatial co-ordinates and other spatial co-ordinates $x^A$ are along the transverse directions to the N=constant surface. Such a co-ordinate change is valid at least locally. The components of the acceleration along N is given by $a^N =a^\mu \partial_\mu N =N a^2$.
 The metric components in this set of co-ordinates are 
 \beq
 g^{NN}=\gamma^{ab} \partial_\mu N \partial _\nu N=N^2a^2, \quad g^{NA}=Na^A
 \eeq
 
 The metric line element now reads:
 \beq
 ds^2 = N^2 dt^2 - \frac{dN^2}{(Na)^2}- d\Omega^2_{\perp} 
 \eeq
 where $d\Sigma^2$ is the line element on the transverse surface.
 
  The unit vector normal to the constant N surface as we calculated is given by $n_a$ and the component of the acceleration along this vector becomes the surface gravity of the horizon $\kappa$ at the N=0 surface i.e in the the limit $N \rightarrow 0$,$Na \rightarrow \kappa$. Therefore, in the limit  of going towards the horizon $N \rightarrow 0$, the metric reads:
  \beq
  ds^2 =N^2 dt^2 - \frac{dN^2}{\kappa^2} - d\Omega_{\perp}^2
  \eeq
  Finally to switch to Rindler-like metric the transformation is $d \ell = dN/a$, $\ell \approx N^2/(2\kappa)$: 
 \beq
 ds^2 = 2\kappa \ell dt^2 -\frac{d \ell^2}{2 \kappa \ell } + d\Omega_{\perp}^2
 \eeq
  We have shown that a general static spacetime with a horizon  can be approximated to  the Rindler spacetime in the limit of going towards horizon. This could be extended to stationary spacetime like the Kerr too but with much more complicated analysis.

\section*{Appendix B: Rindler Hamiltonian in condensed matter}
 
In the following we give an overview of the occurrence of Rindler Hamiltonian in the discussion of bulk-boundary correspondence of topological phases. The discussion in this section is more in the spirit of survey of previous works and is at a qualitative level. Our intention is motivate the idea that the Rindler Hamiltonian is not an exotic object restricted to the context of Hawking-Unruh effect but has wider relevance in condensed matter setting.  The following two sections can be skipped without losing the flow of key ideas in the paper.

The bulk-edge correspondence is one of the fundamental concepts in our understanding of topological phases. A specific way of characterising this correspondence using the ground state entanglement spectrum was initiated by Li and Haldane \cite{LiHaldane} and hinted at in Ref.[\onlinecite{Kitaev06}]. There have been numerous works  proving it using various methods \cite{Fidkowski10,Chandran11,Qi12,Swingle12}. The key idea is that the entanglement spectrum contains with in it key properties of the topological phase. The ``low-lying'' part of the spectrum, for example, directly corresponds to the properties of the edge states at the boundary of a topological phase. Here we shall focus on the approach taken in works Swingle and Senthil \cite{Swingle12}  and Hughes et al \cite{Hughes16}, where they have made use of the Rindler Hamiltonian to calculate the entanglement spectrum and corresponding edge states. The key focus is on the correspondence between 'virtual edge states' at an entanglement cut in the bulk and the physical edge states at the boundary of the system. These 'virtual edge states' correspond to the gapless part of the entanglement spectrum. Both these works consider Lorentz invariant theories, though arguments are given for validity of the results even in the presence of Lorentz symmetry violations. Consider a time slice of a system on x-y plane. Suppose the system is characterised by some quantum hall phase present in the region $x<0$. Then the boundary of the system at $x=0$ has a chiral edge mode. Imagine an entanglement cut through the system given by $y=0$. The entanglement cut divides the system into right and left Rindler wedges. In order to calculate the entanglement properties of the quantum Hall state one considers a Chern-Simons theory in the bulk as the low-energy effective description and integrate out the left half of the system to obtain the reduce density matrix. One could also consider a simpler case of Dirac fermions on the right Rindler wedge in order to compute the entanglement spectrum \cite{Swingle12}. The Dirac equation can be solved with a domain wall profile for the mass term in the x-direction. On solving the equation one finds a zero-mode at the edge of the quantum hall phase which describes a chiral mode. On the other hand the Dirac equation can be solved in the bulk but imposing a "brick-wall" boundary condition near the entanglement cut to obtain the 'virtual' edge mode at the cut \cite{Hughes16}. What this demonstrates is that there is 'virtual' edge state in the entanglement spectrum of the Dirac fermion on a Rindler wedge, signaling its relation to the physical edge mode at the boundary of the system. 

A deeper understanding of this was obtained by computing the gravitational anomalies of the bulk and the edge theories when coupled to a gravtitatonal background \cite{Hughes16,Iqbal16}. The Chiral edge mode is described by a 1+1 dimensional conformal field theory(CFT). It is known that such a CFT exhibits a gravitational anomaly when coupled to a background metric. Now suppose one partitions the system as described above but deform the half-space in the Rindler wedge through a local boost. The expectation is that the entanglement entropy computed using that partition is preserved as the domain of dependence is preserved under such a boost. But coupling to the back ground metric leads to an anomalous change in the entanglement entropy of the edge theory. it was shown that this anomaly is in fact canceled from another gravitational anomaly arising from the 'virtual' edge mode at the entanglement cut and this happens through an anomaly inflow mechanism. A crucial relation to our current work comes in the way such a 'boosting' of the chiral edge theory is realised in a Chern insulator for example. It is realised through changing the velocity of the chiral mode at the edge theory, which also amounts to changing one of the parameters in the Chern insulator Hamiltonian.  Applying potential to the quantum hall system essentially modifies the 'drift velocity' of the edge state and this fact was used to create an analogue of Hawking radiation in a quantum Hall system \cite{Stone13}.



\bibliography{main}

\end{document}